\documentclass[preprint,aps]{revtex4}
\usepackage{mathrsfs}
\usepackage{graphicx}
\usepackage{array,threeparttable}

\newcommand{\PreserveBackslash}[1]{\let\temp=\\#1\let\\=\temp}
\newcolumntype{C}[1]{>{\PreserveBackslash\centering}p{#1}}
\newcolumntype{R}[1]{>{\PreserveBackslash\raggedleft}p{#1}}
\newcolumntype{L}[1]{>{\PreserveBackslash\raggedright}p{#1}}

\begin{document}

\title{Structural, Magnetic and Electronic Properties of the Iron-Chalcogenide A$_x$Fe$_{2-y}$Se$_2$ (A=K, Cs, Rb, Tl and etc.) Superconductors}

\author{Daixiang Mou, Lin Zhao and Xingjiang Zhou$^*$}

\affiliation{
\\$^{1}$National Laboratory for Superconductivity, Beijing National Laboratory for Condensed
Matter Physics, Institute of Physics, Chinese Academy of Sciences,
Beijing 100190, China }
\date{November 20, 2011}

\begin{abstract}

The latest discovery of a new iron-chalcogenide superconductor A$_x$Fe$_{2-y}$Se$_2$(A=K, Cs, Rb, Tl and etc.) has attracted much attention due to a number of its unique characteristics, such as the possible insulating state of the parent compound, the existence of Fe-vacancy and its ordering, a new form of magnetic structure and its interplay with superconductivity, and the peculiar electronic structures that are distinct from other Fe-based superconductors.  In this paper,  we present a brief review on the structural,  magnetic and electronic  properties of this new superconductor, with an emphasis on the electronic structure and superconducting gap. Issues and future perspectives are discussed at the end of the paper.

\end{abstract}


\maketitle

\tableofcontents

\section{Introduction}

The discovery of superconductivity in iron-based compounds in 2008 by Hosono and collaborators\cite{Kamihara} with a superconducting critical temperature up to $\sim$56 K\cite{XHChen,NLWang,ZARenSm,ZAXu} ushered in a second class of ``high temperature superconductors"  after the discovery of the first class of high temperature superconductors in copper-oxide compounds (cuprates) in 1986\cite{Bednorz}.  So far, four main families of Fe-based superconductors have been found, denoted as  ``1111"-type ReFeAsO(Re = rare earth)(FeAs1111)\cite{Kamihara,XHChen,NLWang,ZARenSm,ZAXu}, ``122"-type BFe$_2$As$_2$(B=Ba, Sr, or Ca)(FeAs122)\cite{RotterSC}, ``111"-type AFeAs(A = alkali metal)(FeAs111)\cite{XCWang}, and ``11"-type tetragonal $\alpha$-FeSe(Te)(FeCh11)\cite{FCHsu}.  It is instructive to compare and contrast the Fe-based superconductors to the copper-oxide superconductors in order to pinpoint some essential ingredients in realizing high temperature superconductivity\cite{Ishida,Paglione,Johnston,FWangReview}. There are similarities between the Fe-based superconductors and cuprate superconductors: (1). Structurally speaking, both the Fe-based compounds and cuprates have layered structures. The Fe-based compounds consist of a common FePn(Pn=As or Se) layers which are considered to be essential for the occurrence of superconductivity, similar to the CuO$_2$ planes in cuprates. (2). Superconductivity in the Fe-based compounds is realized in a vicinity of antiferromagnetic state, a case that is similar to that in the cuprates. There are also some significant differences between the Fe-based and copper-based superconductors: (1). Usually the perfectness of CuO$_2$ plane is critical in sustaining superconductivity in cuprate superconductors. But the FeAs(Se) layers appear to be more tolerant to modifications of external perturbations. In fact, substitution of Fe by other ions like Co or Ni can even transform a non-superconductor into a superconductor\cite{ASSefat,ZAXuCoDoping}. (2). It is known that the parent compounds of cuprate superconductors are insulating Mott insulators\cite{PALee}, but most parent compounds of the Fe-based superconductors are bad metals.  (3). In the cuprate superconductors, the electronic structures are mainly dictated by a single Cu 3d$_{x^{2}-y^{2}}$ band. But in the Fe-based compounds, all the five Fe 3d orbitals contribute to the formation of low-lying electronic states, thus forming a typical multi-band system.

The Fe-based superconductors discovered so far possess some common characteristics of electronic structure\cite{DJSingh,KKuroki,FJMa,XJZhou}.  The low-lying electronic excitations are mainly dominated by five 3d orbitals which give rise to a couple of hole-like bands near the zone center $\Gamma$ and electron-like bands near the zone corner {\it M}. Since the parent compound of the Fe-based superconductors is a bad metal, the electron correlation is believed to be not as strong as that in cuprates. It remains under debate whether electrons in the Fe-based superconductors should be treated locally or itenerantly\cite{Ishida,Paglione,Johnston,FWangReview}. Moreover, while it is well-established that in the cuprate superconductors, the superconducting order parameter has predominantly d-wave symmetry, the pairing symmetry in the Fe-based superconductors remains unclear. It has been proposed that the interband scattering between the hole-like bands near $\Gamma$ and the electron-like bands near {\it M} gives rise to electron pairing and superconductivity\cite{KKuroki,Nesting}. On the other hand, another approach based on strong coupling suggested that the parent compound antiferromagnetism could be understood with frustrated Heisenberg model and the local spin coupling could give rise to superconductivity\cite{QMSi,LocalM}. Both of these two approaches give an s$\pm$ pairing symmetry (nodeless gap with sign change between hole and electron Fermi surface sheets). In addition, the orbital degree of freedom has also been proposed\cite{OrbitalFeAs} to play an important role in this multiband system. Enhanced by electron-phonon coupling, orbital order is considered to be the driving force of antiferromagnetic transition and its fluctuation could lead to superconductivity with s++ pairing symmetry (no gap sign change between two kinds of Fermi surface sheets). Although nodeless superconducting gap was revealed by Angle Resolved Photoemission Spectroscopy (ARPES)\cite{LZhaoFeAs,HDing} and gap sign change implied by some experimental techniques\cite{ADChristianson,THanaguri}, at this stage it remains to be investigated which of these candidates provide the best description of the Fe-based compounds.

Very recently, a new superconductor A$_{x}$Fe$_{2-y}$Se$_2$ (A=K, Cs, Rb, Tl and etc.) with T$_c$ around 30 K was reported\cite{JGuo,AKMaziopa,CHLi,MHFang,HDWang}.   This new superconductor triggered a new wave of excitement in the superconductivity community because it exhibits a couple of unique characteristics\cite{IImazinPhys}.  First, while the parent compounds of other Fe-based superconductors are bad metals, it is suggested that the parent compound of this superconductor could be an insulator\cite{MHFang}. Second, while the Fe-sites in FeAs(Se) layers of other Fe-based superconductors are filled, there could be Fe vacancies in this new superconductors\cite{MHFang}. The superconductor may show unique magnetic structure with high magnetic transition temperature and large magnetic moment on the Fe site\cite{WBaoCPL,MWangPRB}. Particularly, the electronic structure of the new superconductor is distinct from other Fe-based superconductors in that no hole-like Fermi Surface around $\Gamma$ is present\cite{TQian,YZhang,DMou,LZhao,XWang}. These characteristics will provide new perspectives on understanding the Fe-based superconductors. On the other hand, many issues remain unclear at this moment that need further experimental and theoretical efforts\cite{Ivanovski}.

In this paper, we will present a brief review on the current status of research on the A$_x$Fe$_{2-y}$Se$_2$ superconductors, with an emphasis on their unique electron structure.  The paper is organized as follows: in Section 2, structural properties including Fe vacancy order and several proposed phase diagrams are first discussed. Then we summarize magnetic structures of several reported phases and their spin dynamics. In Section 3,  electronic properties,  including band structure and Fermi surface, gap structures and pairing symmetry are discussed. In Section 4, we provide a  summary and discussions on the future issues.

\section{Structural and Magnetic Properties}

\subsection{Crystal Structure and Phase Diagram}

The superconductivity with a T$_c$ at $\sim$30 K was first reported in a compound with a nominal chemical formula K$_x$Fe$_2$Se$_2$, which was considered to be isostructural to FeAs122, as shown in Fig.1\cite{JGuo}. Later on Fang \emph{et al}. \cite{MHFang} pointed out that Fe deficiency exists in (Tl$_{1-y}$K$_{y}$)Fe$_x$Se$_2$ compounds, as those discovered previously in the TlFe$_x$Ch$_2$ (Ch=S, Se) compounds\cite{OrderOld}.  By tuning the Fe content $x$, different phases from an insulator to a superconductor can be obtained (Fig.2a and Fig.2b).  At low Fe content $x$, the compound is an antiferromagnetic insulator with a high N$\acute{e}$el temperature. With increasing $x$, the antiferromagnetism is gradually suppressed and superconductivity emerges around $x=1.7$ with only a small fraction of superconducting volume. Bulk superconductivity can be found when $ x \geq1.78$. Similar insulator-to-superconductor transition is also reported in K$_x$Fe$_{2-y}$Se$_2$ by varying the potassium content, $x$\cite{DMWang}.  A sign of possible T$_c$ at 40 K was reported in the A$_x$Fe$_{2-y}$Se$_2$ (A=K, (Tl,K)) system\cite{MHFang,DMWang} but it remains hard to isolate a pure superconducting phase. T$_c$ of the A$_x$Fe$_{2-y}$Se$_2$ (A=K, (Tl,Rb)) superconductor decreases from 31 K to zero with increasing pressure  and then another phase with a T$_c$ as high as 48 K was reported under higher pressure\cite{LSun}.

It is apparent that the physical properties of A$_{x}$Fe$_{2-y}$Se$_2$ rely on both the content of A ($x$) and the content of Fe ($y$). It is therefore essential to have a precise determination of the sample composition in order to build a clear correspondence between the composition, structure and physical properties.  A phase diagram was constructed in Rb$_x$Fe$_{2-y}$Se$_2$ based on composition determination and corresponding magnetic, conductivity and specific heat measurements\cite{VTsurkan}. As shown in Fig.2c, superconductivity was discovered with an Fe content between 1.53 and 1.6, while insulating and semiconducting behaviors were observed with Fe contents at $2-y<1.5$ and $2-y>1.6$, respectively. Another phase diagram based on the Fe valence state ($V_{Fe}$) was proposed in K$_x$Fe$_{2-y}$Se$_2$\cite{YJYan} which is divided into three regions (Fig. 2d). These regions show structural and AFM transitions at similar temperatures. But superconductivity appears only in region II with $1.935<V_{Fe}<2.00$ and T$_c$ is nearly independent of $V_{Fe}$. AFM insulating behavior is found on both sides of the superconducting region II, but they show different Fe vacancy ordering. In region I with $V_{Fe} \geq 2.00$, Fe vacancy order with a {\bf q$_2$}=(1/4, 3/4, 0) is observed while in region III with $V_{Fe}<1.935$, the Fe vacancy order has a wave vector of {\bf q$_1$}=(1/5, 3/5,0).  It has been found that all the samples with a chemical formula K$_{1-x}$Fe$_{1.5+x/2}$Se$_2$ are insulators; superconductivity can be obtained only by adding extra Fe content if keeping the potassium content $x$ at 0.8. Therefore the stoichiometric A$_{0.8}$Fe$_{1.6}$Se$_2$ compound was proposed to be the parent compound of the A$_{x}$Fe$_{2-y}$Se$_2$ superconductors which is an AFM insulator\cite{MHFang,YJYan}.

It is worth noting that the A$_{x}$Fe$_{2-y}$Se$_2$ compounds show peculiar resistivity-temperature dependence.  As shown in Fig.1b, in addition to an abrupt superconducting transition at 31 K,  the resistivity of K$_{0.8}$Fe$_2$Se$_2$ superconductor exhibits a broad hump around 140$\sim$150 K(T$_H$) where there appears to be an insulator-metal transition\cite{JGuo}.  Such a hump is common in other A$_x$Fe$_{2-y}$Se$_2$ superconductors\cite{AKMaziopa,CHLi,MHFang}. It was found that the hump maximum temperature (T$_H$) depends sensitively on the type of $A$ atom\cite{RHLiuEPL} and the Fe content\cite{MHFang,DMWang}. It shows little change with the applied magnetic field\cite{JGuoPressure}. Pressure can gradually suppress the magnitude of the hump with a slight increase of T$_H$\cite{JGuoPressure,YKawasaki,JJYing,GSeyfarth}. It changes non-monotonically with isovalent substitution of Se with sulfur in K$_x$Fe$_{2-y}$(Se$_{2-z}$S$_z$)\cite{HLei}. When comparing to the magnetic susceptibility, no corresponding anomaly is found at the temperature of the broad hump, suggesting it may not be a magnetic transition\cite{XLuo}. Moreover, structural analysis reveals no structural transition around T$_H$\cite{JGuoPressure,WBao,WBaoCPL}. It is possible that such a resistivity hump is related to the phase separation, as will be discussed below.

\subsection{Iron Vacancy Order and Phase Separation}

As mentioned before, the stoichiometry of synthesized samples always deviate from the ideal AFe$_2$Se$_2$ due to the restriction of the Fe valence. Iron vacancy would form an ordered state in Fe-deficient A$_{x}$Fe$_{2-y}$Se$_2$ as reported years ago\cite{OrderOld}. Because of the discovery of superconductivity, many new experimental studies have been carried out recently which have revealed different types of iron ordering in A$_{x}$Fe$_{2-y}$Se$_2$ by transmission electron microscope (TEM)\cite{ZWang,YJSong,JQLi,SMKazakov}, neutron scattering\cite{WBaoCPL,WBao,FYe,VYuPomjakushinJPCM,VYuPomjakushinPRB,MWangPRB}, X-ray diffraction (XRD)\cite{PZavalij,JBacsa} and scanning tunneling microscopy (STM)\cite{WLiSTM,PCai}. Up to now, five different phases were found (Fig.3) which include vacancy free phase(Fig.3a), $\sqrt{5}\times\sqrt{5}$ superstructure phase(Fig.3b), $2\times2$ superstructure phase(Fig.3c), $\sqrt{2}\times2\sqrt{2}$ superstructure phase(Fig.3d) and $\sqrt{2}\times\sqrt{2}$ superstructure phase. From the iron ordering pattern, the Fe content of $\sqrt{5}\times\sqrt{5}$ superstructure phase corresponds to 1.6 and its stoichiometry can be written as K$_{0.8}$Fe$_{1.6}$Se$_2$ or K$_2$Fe$_4$Se$_5$(245 phase) if undoped. Similarly,  both stoichiometries of $2\times2$ and $\sqrt{2}\times2\sqrt{2}$ superstructure phases can be written as K$_2$Fe$_3$Se$_4$(234 phase I and 234 phase II respectively). Table I summarizes observed superstructures in different A$_x$Fe$_{2-y}$Se$_2$ samples.  The vacancy free phase has been observed  only in superconducting samples while the $\sqrt{2}\times2\sqrt{2}$ superstructure is observed only in insulators. The other three superstructures were observed both in superconducting and insulating samples. Compared with other phases that are related to iron vacancy, the origin of the $\sqrt{2}\times\sqrt{2}$ superstructure has not reached a consensus. It was observed by TEM\cite{JQLi}, XRD\cite{VYuPomjakushinPRB} and neutron scattering\cite{MWangPRB}, suggesting it is related to the lattice. But the STM results suggest it may have a magnetic origin\cite{PCai,WLiSTM}.

Temperature dependent neutron scattering studies on  K$_x$Fe$_{2-x/2}$Se$_2$ compounds with various iron content were reported by Bao \emph{et al}. and  a phase diagram was proposed as shown in Fig. 4\cite{WBao}. At high temperature, all samples show tetragonal ThCr$_2$Si$_2$ structure with I4/mmm group symmetry and both K and Fe sites are partially occupied in random. As temperature goes down, for insulator compounds with low iron content, two vacancy ordered state with I4/m($\sqrt{5}\times\sqrt{5}$ superstructure) and Pmna($\sqrt{2}\times2\sqrt{2}$ superstructure) group symmetry appear and coexist with I4/mmm phase between 295 K and 500 K. Only $\sqrt{5}\times\sqrt{5}$ superstructure phase exists when temperature is further lowered below 295 K. For superconductors with higher iron content,  vacancy ordered state with I4/m group symmetry develops after the structure transition at 500 K and maintains to the lowest temperature even below T$_c$, indicating the superconductivity may coexist with the $\sqrt{5}\times\sqrt{5}$ superstructure.

So far it is hard to prepare single phase A$_{x}$Fe$_{2-y}$Se$_2$ samples, especially superconducting ones. For superconducting samples, it is common to observe two sets of {\it c} lattice constants, even for single crystal samples\cite{HDWang,XLuo,HLeiPRB}. The coexistence of multiple phases has been observed in TEM\cite{ZWang,SMKazakov} and STM\cite{PCai,WLiSTM} results(see Fig.5).  While phase separation is possible along the a-b plane, it is interesting to see that, in the superconducting samples, TEM revealed that the Fe-vacancy disorder state (DOS) and order state (OS) alternate along the {\it c}-axis direction (see Fig.5a).  The temperature evolution of the phase separation was also investigated by XRD studies on a K$_{0.8}$Fe$_{1.6}$Se$_2$ superconductor\cite{ARicciSST}. At high temperature above 600 K, no superstructure was found and it is a vacancy-disordered tetragonal phase. After formation of  $\sqrt{5}\times\sqrt{5}$ superlattice at 580 K, the (220) peak splits and another set of superstructure spots appears at 520K, which can be assigned to $\sqrt{2}\times\sqrt{2}$ superlattice. Using nanofocused XRD on different parts of the sample, the proportion of these two phases is found to vary from one part to another\cite{ARicciPRB}. It remains to see whether a pure A$_{x}$Fe$_{2-y}$Se$_2$ superconducting sample can be prepared or  the phase separation is an intrinsic process for superconducting samples.

\subsection{Magnetic Structures}

It has been well-established that the parent compounds of FeAs1111, FeAs122 and FeAs111 have collinear magnetic structure while the FeCh11 phase (FeTe) has a bi-collinear magnetic structure\cite{CCruz,QHuang,SLi,WBao11,MDLumsden}. For the A$_x$Fe$_{2-y}$Se$_2$ compound, the magnetic structures with different iron vacancy ordering are predicted from the theoretical calculations, as summarized in Fig.3.  For the hypothetical stoichiometric AFe$_2$Se$_2$ compound\cite{XYYanPRBVF}, the ground state shows a bi-collinear antiferromagnetic order, with the Fe moments having collinear antiferromagnetic order in each bipartite sublattice (Fig.3a). It results from Se $4p$ orbitals mediated super-exchange interactions of Fe moments, similar to the FeCh11 compound\cite{WBao11}. For the 245 phase with the $\sqrt{5}\times\sqrt{5}$ superstructure\cite{CCaoPRL,XYYanPRBVO,RYuPRB,WLv}, the ground magnetic state favors block AFM structure with all the magnetic moment parallel to {\it c} axis and four nearest iron atoms forming a cluster(Fig.3b). It is ferromagnetic(FM) within one cluster and antiferromagnetic  between the adjacent clusters. Two different magnetic structures were predicted in the compound with the same 234 stoichiometry. For the square ordered 234 phase I\cite{XYYanPRL}, the magnetic structure is {\it c}-AFM order in which the next-nearest Fe moments are  ordered in antiparallel if one just ignores the iron vacancy(Fig.3c).  For the rhombus ordered 234 phase II\cite{XYYanPRL}, it has an A-collinear AFM order magnetic structure where the Fe moments are antiferromagnetically ordered along the line without vacancies(Fig.3d).

Block AFM structure of the 245 phase with the $\sqrt{5}\times\sqrt{5}$ superstructure was established from neutron diffraction on both polycrystalline samples\cite{WBaoCPL,VYuPomjakushinJPCM,VYuPomjakushinPRB} and single crystal samples\cite{MWangPRB,FYe}.  Fig. 6 shows the development of the block AFM structure of the 245 phase in various A$_x$Fe$_{2-y}$Se$_2$ (A=K, Rb, Cs, (Tl,K), (Tl,Rb)) samples\cite{FYe}. The antiferromagnetic N$\acute{e}$el temperature(T$_N$), structural transition temperature(T$_S$) and magnetic moment of different A$_x$Fe$_{2-y}$Se$_2$ samples measured by different experimental techniques are summarized in Table II.  Because the establishment of the block AFM is based on the formation of the $\sqrt{5}\times\sqrt{5}$ superstructure, T$_N$ is always slightly lower than T$_S$, except for a K$_x$Fe$_{2-y}$Se$_2$ insulator where the two transitions occur at nearly the same temperature\cite{WBao}. The A$_x$Fe$_{2-y}$Se$_2$  compounds have a rather high  N$\acute{e}$el temperature (above 500 K),  and large magnetic moment (up to 3$\mu_B$/Fe) which is the largest among all Fe-based compounds.

As mentioned above, due to the existence of multiple phases in A$_x$Fe$_{2-y}$Se$_2$ samples, it remains unclear which phase is really superconducting. A related issue is whether  antiferromagnetism and superconductivity can coexist in the system. On the one hand, coexistence of superconductivity and magnetism in A$_x$Fe$_{2-y}$Se$_2$ was proposed from results of neutron scattering\cite{WBaoCPL,FYe,VYuPomjakushinJPCM,VYuPomjakushinPRB,MWangPRB}, transport measurements\cite{RHLiuEPL,YJYan,VTsurkan}, $\mu$SR\cite{ZShermadini}, M$\ddot{o}$ssbauer\cite{DHRyan} and Raman scattering\cite{AMZhangM}.  As shown in the inset of Fig. 6, the magnetic Bragg intensity from the $\sqrt{5}\times\sqrt{5}$ superstructure shows a kink around T$_c$, indicating that the antiferromagnetism and superconductivity are coupled\cite{WBaoCPL,FYe}.  On the other hand, M$\ddot{o}$ssbauer spectroscopy studies on Rb$_{0.8}$Fe$_{1.6}$Se$_2$ superconductor revealed the presence of 88$\%$ magnetic and 12$\%$ non-magnetic Fe$^{2+}$ species which could be attributed to the $\sqrt{5}\times\sqrt{5}$ superstructure phase and vacancy free phase, respectively\cite{VKsenofontov}. STM results also indicate that the superconductivity comes from the vacancy free phase and the $\sqrt{5}\times\sqrt{5}$ superstructure phase is an insulator(Fig.b-d)\cite{WLiSTM,PCai}.
Further work are needed to reconcile these seemingly conflicting results. Recently,  Hu \emph{et al.} proposed\cite{WLiCalculation} that the vacancy order free phase(also superconducting phase) may have block-AFM ground state as well, similar to that in $\sqrt{5}\times\sqrt{5}$ superstructure phase. Since these two block-AFM states can be described by the same magnetic model, two separated phases can couple with each other in one sample.

\subsection{Spin Dynamics}

The spin dynamics of the insulating Rb$_{0.89}$Fe$_{1.58}$Se$_2$ compound with a $\sqrt{5}\times\sqrt{5}$ superstructure has been investigated by inelastic neutron scattering(INS)\cite{MWangIN}. As shown in Fig. 7, the spin waves exist in three separated energy ranges. The lowest branch, which is an acoustic mode arising mostly from antiferromagnetic interactions of the FM blocked spins\cite{RYuPRB,YZhouPRB}, starts from 9 meV to 70 meV. The other two branches, which are optical spin waves associated with exchange interactions of iron spins within the FM blocks, are from 80 meV to 140 meV and 180 meV to 230 meV respectively. The magnetic exchange couplings, obtained by fitting the data with effective $J_1-J_1'-J_2-J_2' -J_3-J_c$ Heisenberg model, are summarized in Table III, together with those in Fe$_{1.05}$Te\cite{OJLipscombe} and CaFe$_2$As$_2$\cite{JZhao} for comparison. Although their static antiferromagnetic orders have completely different structures, these three iron-based compounds have comparable effective exchange couplings $J_2$, which is mainly determined by a local superexchange mediated by As or Se/Te\cite{QMSi}. This is consistent with the idea that $J_2$ is the leading parameter of ground magnetic state and closely related to superconductivity in the Fe-based superconductors\cite{JPHuArxiv}.

It is predicted that, if there is a sign change in the superconducting order parameter, a spin resonance mode with an energy between one and two times of the superconducting gap would appear at the wave vector connecting two parts of the Fermi surface with opposite gap signs\cite{PMonthoux}. The spin resonance mode,  generally taken as a hallmark of unconventional pairing symmetry of superconductivity, has been observed in cuperate superconductors\cite{ResonanceCu,MEschrig}, heavy Fermion superconductors\cite{ResonanceHF} and the Fe-based superconductors\cite{ADChristianson,MDLumsden}. Such a resonance mode has also been revealed in Rb$_{0.8}$Fe$_{1.6}$Se$_2$ superconductor\cite{JTPark}. As shown in Fig. 8, the intensity is obviously enhanced at 14 meV across T$_c$ at a wave vector (0.5, 0.3125, 0.5), in contrast to (0.5, 0.5, 0.5) where the resonance has been theoretically predicted in d-wave pairing symmetry\cite{TAMailer,TDas} or (0.5, 0, 0.5) where it is usually found in other Fe-based superconductors\cite{ADChristianson,MDLumsden}. Temperature dependence of the measured resonance intensity follows an order-parameter-like increase below T$_c$, indicating it is related to the superconducting transition (Fig.8c). The ratio of  $\hbar\omega_{Res}/k_BT_c$ is 5.1$\pm$0.4, slightly above the nearly universal ratio of 4.3 estimated for 122-compounds, but is close to that in FeTe$_{1-x}$Se$_x$, LiFeAs, La-1111, and cuprate superconductors\cite{DSInosov,GYu,MDLumsden,Paglione}. Another scaling parameter 2$\Delta/\hbar\omega_{Res}$  would be 0.7$\pm$0.1, if taking the superconducting gap as 10 meV measured by ARPES\cite{YZhang,DMou,XWang,LZhao}. It is slightly larger than the typical value of 0.64 for cuprates but smaller than 1, following the general trend found in all Fe-based superconductors\cite{DSInosov,GYu}.

It is expected that antiferromagnetic spin fluctuations, which is proposed to be a strong candidate of the pairing glue in the Fe-based superconductors and has been observed in 11-type iron-chalcogenide and iron-pnictide superconductors, would introduce a Curie-Weiss upturn near T$_c$  in 1/(T$_1$T) in the nuclear magnetic resonance measurements\cite{TImai,FLNing}.  However,  no indication of such an enhancement is detected in several A$_x$Fe$_{2-y}$Se$_2$(A=K, (Tl,Rb)) superconductors so far\cite{WYu,DATorchetti,HKotegawa,LMaPRB}. Instead, the results are more similar to the overdoped non-supercondcting 122-type iron-pnictide superconductors where the absence of Curie-Weiss upturn was interpreted as due to the lack of nesting between the hole and electron Fermi surface sheets because the hole band sinks below the Fermi level by electron doping. Similar band structure was observed by ARPES in various A$_x$Fe$_{2-y}$Se$_2$ superconductors, as described below.

\section{Electronic Properties}

\subsection{Band Structure Calculations}

In A$_x$Fe$_{2-y}$Se$_2$, the intercalated $A$ can donate electrons into the Fe$_{2-y}$Se$_2$ layers. The electronic structure of the iron vacancy free phase, A$_x$Fe$_2$Se$_2$ (A=K, Cs, Rb, Tl), can be treated as electron-doped FeSe (doping level is $x$) with the chemical potential raising up,  or hole-doped AFe$_2$Se$_2$ (doping level $1-x$) with the chemical potential moving down, as confirmed by LDA calculations\cite{IRShein,CCaoCPL,IANekrasov,XYYanPRBVF}.  The calculated electronic structures have little dependence on the $A$ element because the density of states (DOS) around the Fermi level are mainly associated with Fe-$3d$ and Se-$4p$ orbitals\cite{CCaoCPL}. The calculated band structure of  AFe$_2$Se$_2$ (A=K, Cs) is shown in Fig. 9b, together with that of BaFe$_2$As$_2$ for comparison\cite{IANekrasov}.  The band structure of A-deficient A$_x$Fe$_2$Se$_2$ looks similar to that of BaFe$_2$As$_2$ if the chemical potential of  AFe$_2$Se$_2$ band is lowered, as shown in Fig. 9b.  Compared to BaFe$_2$As$_2$ where both the two e$_g$ orbitals ($3z^2-r^2$ and $x^2-y^2$) and the three t$_{2g}$ orbitals ($xz$, $yz$ and $xy$ orbitals) are at the Fermi level, the e$_g$ states in A$_x$Fe$_2$Se$_2$ are sinked below E$_F$ and the states at E$_F$ are only contributed by the t$_{2g}$ states. As a result, the Fermi surface of AFe$_2$Se$_2$(Fig. 9c) consists of two quasi-two-dimensional electron-like Fermi surface sheets near $X$  and a small electron-like Fermi pocket near $Z$ which is more three-dimensional.  Note that in this case all the Fermi surface sheets are electron-like.

The electronic structure of various Fe-deficient A$_x$Fe$_{2-y}$Se$_2$ phases has also been calculated, such as vacancy free phase with bi-collinear AFM\cite{XYYanPRBVF}, 245 phase with block AFM\cite{CCaoPRL,XYYanPRBVO}, 245 phase in non-magenetic state\cite{CCaoPRL,XYYanPRBVO}, and 234 phase II with A-collinear AFM\cite{XYYanPRL}.  Fig. 10 shows band structure of two typical Fe-vacancy-ordered phases. Of particular interest is the electronic structure of the A$_{0.8}$Fe$_{1.6}$Se$_2$ (245 phase) that has been commonly revealed in structural analysis.  The calculated band structure for the 245 phase with block AFM (Fig. 10a) indicates that it is a semiconductor with band gap as large as 400-600meV\cite{CCaoPRL,XYYanPRBVO}.  This is totally different from the vacancy free A$_x$Fe$_{2}$Se$_2$ phase or any other Fe-based superconductors.  Detailed analysis of the orbital contribution indicates that the top valence bands are composed of both Fe-$3d$ and Se-$4p$ orbitals and the lower conduction bands are mainly contributed by the Fe-$3d$ orbital.  All the five up-spin orbitals are almost fully filled, while the down-spin orbitals are partially filled by $d_{z^2}$ , $d_{xy}$ , and $d_{yz}$ orbitals, suggesting the Hund coupling is larger than the crystal-field splitting induced by Se atoms. The 234 phase II in magnetic ground state was also indicated by LDA calculations as an semiconductor with a smaller band gap\cite{XYYanPRL}, as shown in Fig. 10b. It varies from tens to more than a hundred meV between different compounds.  It has been predicted that Mott insulator behavior can exist  in both the 234 and 245 vacancy-ordered phases\cite{RYuPRL,CCaoPRB,YZhouEPL,LCraco}. The iron vacancy is expected to reduce the neighboring iron number in 234 or 245 phases, thus making the electron hoping term between neighboring irons smaller and then the bandwidth narrower. This reduction of bandwidth tends to push the vacancy ordered phase of  A$_x$Fe$_{2-y}$Se$_2$ into the Mott insulator regime.

\subsection{Band Structure and Fermi Surface}

Angle-resolved photoemission (ARPES) experiments have been successfully carried out on the A$_x$Fe$_{2-y}$Se$_2$(A=K, Cs, (Tl,K), (Tl, Rb)) compounds\cite{TQian,YZhang,DMou,XWang,LZhao,FChen,DMouUn}. Fig.11a shows a wide energy scan of a photoemission spectrum which includes shallow core levels and the valence band of the K$_{0.8}$Fe$_{1.7}$Se$_2$ superconductor\cite{TQian}. Besides a sharp peak of K $3p$ core level at a binding energy of 17.55 eV and another peak  at 12 eV, a weak feature at 0.9 eV is also revealed. A weakly-dispersive broad band is observed around this energy from the ARPES measurements (Fig.11c and 11d)\cite{TQian}. This feature shifts to lower binding energy with increasing temperature\cite{XWang}. By systematically analyzing the electronic structures of  superconducting, semiconducting and insulating K$_x$Fe$_{2-y}$Se$_2$ samples, it is suggested that it may come from an insulating phase in the samples\cite{FChen}.

Detailed band structure in the vicinity of Fermi level in several A$_x$Fe$_{2-y}$Se$_2$ superconductors have been investigated and typical results are shown in Fig. 12.  The band structure of (Tl,Rb)$_x$Fe$_{2-y}$Se$_2$ superconductor (T$_c$=32 K) around $\Gamma$ shows two electron bands  ($\alpha$ and $\beta$ in Fig. 12a) with the $\alpha$ band bottom barely touching the Fermi level. There is also a hole-like band sinking below the Fermi level (Fig. 12b)\cite{DMou}.  In (Tl,K)$_x$Fe$_{2-y}$Se$_2$ and K$_x$Fe$_{2-y}$Se$_2$ superconductors\cite{XWang,LZhao}, in additional to two electron bands, two hole bands(denoted as GA and GB in Fig.12j) below the Fermi level are observed.   Around $M$, one electron-like band is clearly observed (Fig. 12c and 12d).  But a more detailed measurements (Fig. 12e and 12f) indicate that the band is composed of two bands with similar Fermi momenta but with different band bottoms, one at $\sim$ 40 meV and the other at $\sim$60 meV (Fig. 12f). The k$_Z$ dependence of these bands are obtained by ARPES using different photon energies\cite{YZhang}. As shown in Fig.13, the band crossings around $M$ do not change much at different k$_Z$, indicative of their quasi-two-dimensional nature. However, the electron-like $\alpha$ band near $\Gamma$ exhibits obvious k$_Z$ dependence: an electron-like band crosses the Fermi level near $Z$ but it is above the Fermi level near $\Gamma$\cite{YZhang}.

The A$_x$Fe$_{2-y}$Se$_2$ superconducting compounds exhibit similar Fermi surface topology, as summarized in  Fig. 14.  They all show two electron-like pockets around $\Gamma$ (denoted as $\alpha$ and $\beta$) and one electron-like pocket  around $M$ ($\gamma$) which is in fact composed of two nearly degenerate Fermi surface sheets.   Around $\Gamma$ point, no Fermi surface sheet or only one small sheet were first reported in K$_x$Fe$_{2-y}$Se$_2$ and Cs$_x$Fe$_{2-y}$Se$_2$\cite{TQian,YZhang}. Later results on (Tl,Rb)$_x$Fe$_{2-y}$Se$_2$ with improved data quality revealed two electron-like Fermi surface sheets around $\Gamma$\cite{DMou}. This distinct Fermi surface topology was further confirmed by other measurements of (Tl,K)$_x$Fe$_{2-y}$Se$_2$ and K$_x$Fe$_{2-y}$Se$_2$\cite{XWang,LZhao}, making it a common Fermi-surface topology in the A$_x$Fe$_{2-y}$Se$_2$ superconductors. By calculating the enclosed area of all measured Fermi pockets, the doping concentration can be estimated to be 0.27 e/Fe for (Tl$_{0.58}$Rb$_{0.42})$Fe$_{1.72}$Se$_2$\cite{DMou} and 0.32 e/Fe for (Tl$_{0.63}$K$_{0.37})$Fe$_{1.78}$Se$_2$\cite{XWang}. It is 0.18 e/Fe for both compounds, if only the two degenerate pockets around M are counted.

It is noted that the measured Fermi surface and band structure of the A$_x$Fe$_{2-y}$Se$_2$ superconductors are qualitatively consistent with band structure calculations on the vacancy free AFe$_2$Se$_2$ phase (Fig. 9). One outstanding difference is the $\beta$ Fermi surface sheet around $\Gamma$ (Fig.13) that is absent in the band calculations. We note here that, since the exact superconducting phase in A$_x$Fe$_{2-y}$Se$_2$ remains unclear, one should be cautious in making such a direct comparison to jump into some conclusions. An immediate issue is on the origin of this $\beta$ Fermi surface.  The first possibility is a surface state. While surface state on some Fe-based compounds like the FeAs1111 system was reported before\cite{HYLiu}, it has not been observed in the ``11"-type Fe(Se,Te) system\cite{YXia}. The second possibility is whether the $\beta$ band can be caused by the folding of the electron-like $\gamma$ Fermi surface near $M$. It is noted that the Fermi surface size, the band dispersion, and the band width of the $\beta$ band at $\Gamma$ is similar to that of the $\gamma$ band near $M$.  A band folding picture would give a reasonable account for such a similarity if there exists a ($\pi$,$\pi$) modulation in the system that can be either structural or magnetic. An obvious issue with this scenario is that, in this case, one should also expect the folding of the $\alpha$ band near $\Gamma$ onto the $M$ point; but such a folding is not observed at the $M$ point (Fig. 12c and Fig. 12d).  The third possibility is whether the measured $\beta$ sheet is a Fermi surface at a special k$_Z$ cut. Due to its weak intensity, this $\beta$ Fermi surface  is not revealed in the k$_Z$ dependence measurements (Fig.13). But almost the same Fermi crossing was observed with three different energies(6.994 eV, 21.2 eV and 40.8 eV), implying it is nearly two-dimensional-like\cite{DMou,DMouUn}.  Note that this $\beta$ Fermi surface should not be confused with the $\alpha$ pocket at $Z$ because they have very different size, particularly they are observed simultaneously near $\Gamma$ point.  The origin of the $\beta$ Fermi surface is still an open question to be further addressed.

It is under debate whether the superconducting phase in A$_x$Fe$_{2-y}$Se$_2$ coexists with the Fe vacancy ordered $\sqrt{5}\times\sqrt{5}$ phase\cite{FYe}. Under such a circumstance, one would expect the Brillouin zone for the $\sqrt{5}\times\sqrt{5}$  phase to be 1/5 of the original 1$\times$1 phase (Fig. 3e). The $\sqrt{5}\times\sqrt{5}$ superstructure would produce multiple folded Fermi surface sheets from the original one pocket near M or near $\Gamma$. However, no indication of such folded Fermi surface sheets are observed in the measured results (Fig. 14). While one cannot fully rule out that the folded Fermi surface is not seen because they may be too weak, their absence is not compatible with the scenario of coexistence of superconductivity and the $\sqrt{5}\times\sqrt{5}$ superstructure. The same argument is also applicable to other possible ordered superstructures like the  $2\sqrt{2}\times\sqrt{2}$ and $2\times2$ phases.

Study on the electronic structure evolution from an insulator to a superconductor is helpful in addressing the key issue on what the parent compound is for the A$_x$Fe$_{2-y}$Se$_2$ superconductors. Valence bands were measured on three types of K$_x$Fe$_{2-y}$Se$_2$ compounds (insulator, semiconductor and superconductor)\cite{FChen}. Comparing with a superconductor, no band was observed within the 0.5 eV energy range below E$_F$  in an antiferromagnetically insulating sample. Two high energy features around 0.7 eV and 1.6 eV binding energies were observed in all three type of samples.  It was found that, by increasing the temperature or reducing the photon intensity, these two features shift toward low binding energy, which is typical behavior called charging effect in photoemission measurements on insulating samples. On the other hand, no such charging effect was observed in low energy feature of a superconductor. Therefore, the high energy features (0.7 eV and 1.6 eV)  and low energy near-E$_F$ features may not come from the same phase in the sample: the former may come from the insulating phase while the latter from metallic or superconducting phase. Similar phase separation behavior was also found in the semiconducting sample. Such results are consisting with the phase separation picture also revealed by TEM\cite{ZWang} and STM\cite{WLiSTM,PCai}.

In contrast to the dramatic difference between the electronic structure of the insulating and superconducting phases, the low energy features of the semiconductor are reminiscent to those in the superconducting sample. As illustrated in Fig. 15b, only hole-like  band with a band gap of $\sim$20 meV was observed around $\Gamma$.  The band structure of the semiconductor seems to resemble that of the superconductor by shifting the chemical potential of the superconductor downwards by 55 meV and coincidentally the chemical potential lies in the gap between the electron and hole bands. No shadow band possibly caused by magnetic order or lattice superstructure was observed in this semiconductor. These observations led the authors\cite{FChen} to propose that the actual parent compound of K$_x$Fe$_{2-y}$Se$_2$ superconductor is not the  insulating phase, but the semiconducting phase; superconductivity can be obtained by doping the semiconductor. More work needs to be done to confirm this interesting scenario.

\subsection{Superconducting Gap and Pairing Symmetry}

ARPES can directly measure superconducting gap and its momentum dependence, thus providing key information on the pairing symmetry of superconductivity. Fig. 16 first shows typical ARPES results on measuring temperature dependence of superconducting gap in (Tl,Rb)$_x$Fe$_{2-y}$Se$_2$  superconductor\cite{DMou,DMouUn}. The photoemission data are taken on different Fermi surface sheets at different temperatures (Fig. 16a and 16e) and the photoemission spectra (energy distribution curves, EDCs) on the Fermi surface are symmetrized (Fig. 16c and 16f) to visualize the superconducting gap opening and obtain the gap size. It is clear that for both the $\beta$ and $\gamma$ Fermi surface sheets the superconducting gap opens right below T$_c$=32 K.  Also the quasiparticle peak sharpens up while entering the superconducting state. As seen in Fig. 16b, a sharp quasiparticle peak with a narrow width of only 9 meV (Full-width-at-half-maximum) is observed  at low temperature for the $\gamma$ Fermi surface near $M$.   The superconducting gap size can be determined by taking the EDC  peak position. As seen in Fig. 16d, the temperature dependence of gap size on the $\gamma$ Fermi surface near $M$ follows the standard BCS form with a  $\Delta_0\sim$9.7 meV\cite{DMouUn}.  Similar temperature dependence of the superconducting gap is also reported in other A$_x$Fe$_{2-y}$Se$_2$ (A=K, (Tl,K)) superconductors, but with different gap sizes\cite{YZhang,XWang,LZhao}. With respect to the small $\alpha$ pocket, no gap opening was detected in (Tl,Rb)$_x$Fe$_{2-y}$Se$_2$ because the band is slightly above the Fermi level near $\Gamma$\cite{DMou}. A superconducting gap of $\sim$7 meV was reported below T$_c$ in K$_x$Fe$_{2-y}$Se$_2$ on the $\alpha$ Fermi surface near $Z$ with a clear Fermi crossing\cite{YZhang}.

The ARPES measurements on different A$_x$Fe$_{2-y}$Se$_2$ (A=K, (Tl,K), (Tl, Rb)) superconductors give nearly isotropic superconducting gap on the $\gamma$ Fermi surface around $M$\cite{YZhang,DMou,XWang,LZhao,GapNote} (Fig. 17e). In order to measure the momentum-dependent superconducting gap on the weak $\beta$ Fermi surface around $\Gamma$, high resolution laser-based ARPES measurements have been performed which also give a nearly isotropic superconducting gap (Fig. 17d)\cite{DMouUn}. It was also found that the gap sizes on both  $\alpha$ and $\gamma$ Fermi surface sheets show little variation at different Fermi crossing along k$_Z$\cite{YZhang}. So far no evidence of gap node is observed in the  A$_x$Fe$_{2-y}$Se$_2$ superconductors.

The low-temperature specific heat measurement\cite{BZheng} and NMR measurement\cite{HKotegawa} also suggest a nodeless gap.  The scanning tunneling spectroscopy measurements, which reveal the averaged electronic structure of the entire Brillion Zone, show a 7 meV superconducting gap in the cleaved K$_x$Fe$_{2-y}$Se$_2$ sample\cite{PCai}. For thin film sample grown by MBE, however,  two superconducting gaps with much smaller size, 1 meV and 4 meV, were reported(Fig.5c)\cite{WLiSTM}. Table IV summarizes the superconducting  gap size of different A$_x$Fe$_{2-y}$Se$_2$ samples measured by different techniques. The ratio of 2$\Delta$/k$_B$T$_c$ obtained from single crystal samples is larger than the traditional BCS weak-coupling value of 3.54, which indicates that A$_x$Fe$_{2-y}$Se$_2$ superconductor is in a strong-coupling regime in the BCS picture. On the other hand, the small ratio of 0.9 and 3.5 obtained in MBE-grown film samples puts it in a weak-coupling regime. This difference between the bulk samples and film samples needs to be further clarified.

In the Fe-based superconductors, it has been proposed that the interband scattering between the hole-like bands near $\Gamma$ and the electron-like bands near $M$ gives rise to electron pairing and superconductivity with a s$\pm$ symmetry (Fig. 18, left panel)\cite{KKuroki,Nesting}.  With the absence of hole-like Fermi surface around $\Gamma$, this picture is no longer applicable in the A$_x$Fe$_{2-y}$Se$_2$ superconductors. In this case, from a weak coupling point of view, the repulsive inter-electron-pocket pair scattering between two $M$ pockets (Fig. 18, right panel) would give a d$_{x^{2}-y^{2}}$ pairing symmetry\cite{KKuroki,FWang, TAMailer, TDas}. On the other hand, in the viewpoint of a doped Mott insulator, by calculating the $t-J$ model, both s-wave and d-wave pairing are possible at different regions of phase diagram\cite{YZhouEPL, RYuArxiv}. But s-wave pairing symmetry is robust if the antiferromagnetic $J_2$ is the main factor for pairing and the $J_1$ is ferromagnetic\cite{CFangPRX}. It is also proposed that orbital order and its fluctuations would give rise to a s++ wave pairing with the presence of a moderate electron-phonon interaction\cite{TSaito}.

The observation of a nearly isotropic, nodeless superconducting gap in A$_x$Fe$_{2-y}$Se$_2$ superconductors favors an s-wave pairing symmetry. It is also considered to be consistent with a nodeless d-wave pairing because the d-wave node is along the diagonal of 1 Fe Brillouin zone (Fig.18, right panel);  the $\gamma$ Fermi surface sheet may show a nearly isotropic gap because it does not cross this node line. However, as pointed out by Mazin\cite{IImazindwave}, if there is a sign change between the (0,-$\pi$) Fermi pocket and ($\pi$,0) pocket, gap node would still be expected at the crossing points of two Fermi pockets around the same $M$ point due to the folding from 1 Fe Brillouin zone to 2 Fe Brillouin zone which is then inconsistent with existing experimental results. Up to now no conclusive answer to the pairing symmetry of the A$_x$Fe$_{2-y}$Se$_2$ superconductors has been reached yet.  We note that, the weak $\beta$ Fermi pocket around $\Gamma$ crosses the node line in d-wave pairing, if it is proven to be intrinsic, its nearly isotropic gap (Fig. 17d)\cite{DMou,DMouUn} could rule out the d-wave pairing scenario.

\subsection{Electron Dynamics}

The coupling between electron and boson modes plays an important role in giving rise to the electron pairing. Such a coupling can be probed directly by ARPES by measuring the electron self-energy which manifests itself as a kink structure on a band dispersion\cite{Damascelli,XJZhouSchriefferbook}. A number of phonon modes are observed in the A$_x$Fe$_{2-y}$Se$_2$ (A = K, Rb, Tl) superconductors by Raman scattering and optical studies\cite{AMZhang,ZGChen}. A magnetic resonance mode has been predicted\cite{TAMailer,TDas} and observed (Fig. 8) in the superconducting state of Rb$_x$Fe$_{2-y}$Se$_2$ at an energy of 14 meV\cite{JTPark}. It is therefore interesting to investigate the electronic dynamics in A$_x$Fe$_{2-y}$Se$_2$ superconductors. Fig. 19 shows the measured band dispersion of the (Tl,Rb)$_x$Fe$_{2-y}$Se$_2$ superconductor around $M$ point at different temperatures, together with the extracted effective imaginary and real parts of the electron self-energy\cite{DMouUn}. The transition near 7 meV in the real part of electron self-energy (Fig. 19c) is caused by the superconducting gap opening and is not the effect of electron-boson coupling.  Except for this feature, no obvious dispersion kink is observed in the real part of self-energy at both normal and superconducting states.  These observations indicate that the electron-boson coupling is weak in the A$_x$Fe$_{2-y}$Se$_2$ superconductors.

\section{Summary and Perspective}

Although it has been only one year since the discovery of the A$_x$Fe$_{2-y}$Se$_2$ superconductors, significant progress has been made in studying their unique structural, magnetic and electronic properties.  Unique characteristics have been revealed and studied, such as the existence of Fe vacancy and its ordering, the novel magnetic structure of block antiferromagnetism with large magnetic moment and high N$\acute{e}$el temperature,  and the possible insulating nature of its parent compound.  In particular, key insights have been drawn in understanding the pairing mechanism of the Fe-based superconductors with its distinct electronic structure, i.e., with the absence of hole-like Fermi surface around $\Gamma$, high T$_c$ ($\sim$30 K at ambient pressure and 48 K under pressure) is achieved in this system.


On the other hand, there are complications and open questions around this new system.  First and foremost, what is the exact superconducting phase in the A$_x$Fe$_{2-y}$Se$_2$ superconductors?   STM and ARPES data favor Fe-vacancy-free A$_x$Fe$_{2}$Se$_2$ phase, but neutron scattering and Raman studies indicate that superconductivity is associated with the Fe-vacancy ordered $\sqrt{5}\times\sqrt{5}$ superstructure phase.  The answer becomes complicated due to the existence of multiple phases in the superconducting samples which leads to another question of phase separation. Is the phase separation in the A$_x$Fe$_{2-y}$Se$_2$ superconductors inevitable or a problem of sample preparation? Can the phase separation properly account for many unusual properties observed so far?  The final answer to the superconducting phase will help in answering the question of its parent compound: whether it is an insulator, or a semiconductor, or a poor metal.  Of course, the ultimately important issue is to reveal the pairing mechanism of superconductivity.  To conclude, the A$_x$Fe$_{2-y}$Se$_2$ superconductors provide a new platform for investigating the properties and mechanism of superconductivity. The storm caused by this new superconducting family is continuing and more exciting results are expected with the future efforts.

We would like to thank collaborations with Shanyu Liu, Xiaowen Jia, Junfeng He, Yingying Peng, Li Yu, Xu Liu, Guodong  Liu, Shaolong He, Xiaoli Dong, Jun Zhang, Hangdong Wang, Chiheng Dong, Minghu Fang, J. B. He, D. M. Wang, G. F. Chen, J. G. Guo, X. L. Chen, Xiaoyang Wang, Qinjun Peng, Zhimin Wang, Shenjin Zhang, Feng Yang, Zuyan Xu, and Chuangtian Chen. This work is financially supported by NSFC (Grant No. 10734120) and the MOST of China (973 program No: 2011CB921703).

$^{*}$Corresponding author: XJZhou@aphy.iphy.ac.cn

\clearpage

\begin{center}
\begin{table}
\resizebox{.8\textwidth}{!}{%
\begin{threeparttable}
\renewcommand\arraystretch{0.8}
\caption{Superstructure phases Reported in A$_x$Fe$_{2-y}$Se$_2$}
\begin{tabular}{|L{0.2\columnwidth}|c|C{0.2\columnwidth}|r|}
 \hline
Superstructure              &SC/Insulator  &Sample                                             &Ref.                   \\
\hline
                                    &        &$K_xFe_{2-y}Se_2$                                    &\cite{ZWang} \cite{WBao} \cite{WBaoCPL} \cite{PZavalij} \cite{JQLi}\\
                                   & SC     &$Rb_yFe_{1.6}Se_2$                                    &\cite{MWangPRB}\\
$\sqrt{5}\times\sqrt{5}$  &         &$Cs_xFe_{2-y}Se_2$                                  &\cite{VYuPomjakushinPRB}\\
                                    \cline{2-4}
                                    &           &$K_xFe_{2-y}Se_2$                                &\cite{ZWang} \cite{JBacsa}\cite{WBao} \cite{WLiSTM} \cite{YJYan}\\
                                    &Insulator &$Rb_yFe_{1.6}Se_2$                                    &\cite{MWangPRB}\\
                                    &           &$K_x(Fe,Co)_{2-y}Se_2$                                 &\cite{SMKazakov}\\
\hline
$2\times2$                     &SC    &$K_xFe_{2-y}Se_2$                                    &\cite{ZWang}\\
                                    \cline{2-4}
                                     &Insulator &$K_xFe_{2-y}Se_2$                                   &\cite{ZWang} \cite{YJYan}\\
\hline
$\sqrt{2}\times2\sqrt{2}$  &Insulator &$K_x(Fe,Co)_{2-y}Se_2$                              &\cite{SMKazakov}\\
\hline
                                   &        &$K_xFe_{2-y}Se_2$                                    &\cite{PCai} \cite{JQLi} \\
                                   & SC     &$Rb_yFe_{1.6}Se_2$                                    &\cite{MWangPRB} \\
$\sqrt{2}\times\sqrt{2}$   &         &$Cs_xFe_{2-y}Se_2$                                  &\cite{VYuPomjakushinPRB}\\
                                    \cline{2-4}
                                    &Insulator &$K_xFe_{2-y}Se_2$                                &\cite{WBao} \cite{YJYan}\\
                                    &       &$K_x(Fe,Co)_{2-y}Se_2$                                 &\cite{SMKazakov}\\
\hline
Vacancy Free                   &SC    &$K_xFe_{2-y}Se_2$                                        &\cite{ZWang} \cite{WLiSTM}\\
\hline
\end{tabular}
\end{threeparttable}}%
\end{table}
\end{center}

\begin{center}
\resizebox{.8\textwidth}{!}{%
\begin{threeparttable}
\caption{Structural transition temperature(T$_S$), magnetic transition temperature(T$_N$) and the ordered magnetic moment of Fe at low temperature(M) and room temperature(m) reported in A$_x$Fe$_{2-y}$Se$_2$.}
\begin{tabular}{L{0.2\columnwidth}C{0.1\columnwidth}C{0.1\columnwidth}C{0.1\columnwidth}C{0.1\columnwidth}C{0.2\columnwidth}r}
\hline \hline
Sample                                      &T$_C$(K)   &T$_S$(K)   &T$_N$(K)  &M($\mu_B$)    &m($\mu_B$)  &Ref. \\ \hline
K$_{0.8}$Fe$_{1.6}$Se$_2$       &32             &578            &559            &3.31               &2.76               &\cite{WBaoCPL}  \\
K$_{0.85}$Fe$_{1.83}$Se$_{2.09}$    &30     &                 &532          &4.7                    &3.0                &\cite{DHRyan}   \\
K$_y$Fe$_{2-x}$Se$_2$               &29.5        &                  &               &                       &2.55               &\cite{VYuPomjakushinJPCM} \\
K$_{0.7}$Fe$_{1.7}$Se$_2$       &31             &560            &               &                        &                     &\cite{JQLi} \\
K$_{0.99}$Fe$_{1.4}$Se$_2$      &insulator             &500            &500         &3.16                  &                     &\cite{WBao} \\ \hline
Rb$_2$Fe$_4$Se$_5$                  &32             &515            &502        &3.3                    &2.95               &\cite{FYe}  \\
 Rb$_y$Fe$_{2-y}$Se$_2$             &31.5       &                   &           &                           &2.15               &\cite{VYuPomjakushinJPCM}   \\  \hline
 Cs$_2$Fe$_4$Se$_5$                 &29             &500            &471        &3.4                    &2.9                  &\cite{FYe}  \\
Cs$_{0.8}$Fe$_2$Se$_{1.96}$     &28.5           &                &477        &                         &                      &\cite{ZShermadini} \\
Cs$_y$Fe$_{2-x}$Se$_2$            &28.5          &500             &             &                       &                        &\cite{VYuPomjakushinPRB} \\
Cs$_{0.8}$(FeSe$_{0.98}$)$_2$    &29.6          &                &478.5      &                      &                        &\cite{ZShermadini} \\ \hline
(Tl,K)$_2$Fe$_4$Se$_5$             &28               &533           &506        &3.2                   &2.6                   &\cite{FYe}  \\ \hline
(Tl,Rb)$_2$Fe$_4$Se$_5$           &32               &512            &511        &3.2                &                         &\cite{FYe}    \\
\hline \hline
\end{tabular}
\end{threeparttable} }%
\end{center}

\begin{center}
\resizebox{.8\textwidth}{!}{%
\begin{threeparttable}
\caption{Effective exchange couplings in three typical iron-based compounds}
\begin{tabular}{L{0.18\columnwidth}C{0.12\columnwidth}C{0.1\columnwidth}C{0.1\columnwidth}C{0.1\columnwidth}C{0.1\columnwidth}C{0.1\columnwidth}C{0.1\columnwidth}r}
\hline \hline
Compond                                      &J$_1$ or J$_{1a}$   &J'$_1$ or J$_{1b}$   &J$_2$ or J$_{2a}$  &J'$_2$ or J$_{2b}$    &J$_3$  &J'$_3$ &J$_c$ &Ref. \\ \hline
Rb$_{0.89}$Fe$_{1.58}$Se$_2$       &-36                  &15                &12                         &16                   &9                   &0    &1.4    &\cite{MWangIN}   \\
Fe$_{1.05}$Te\tnote{a}                    &-17.5            &-51                &21.7                       &21.7                   &6.8                &      &                   &\cite{OJLipscombe} \\
CaFe$_2$As$_2$\tnote{b}                 &50                 &-5.7               &19                         &                          &                  &      &         &\cite{JZhao} \\
\hline \hline
\end{tabular}
\begin{tablenotes}\small
\item[a]Data fitted with $J_1-J'_1-J_2-J'_2 -J_3$ Heisenberg model.
\item[b]Data fitted with $J_{1a}-J_{1b}-J_2$ Heisenberg model.
\end{tablenotes}
\end{threeparttable} }%
\end{center}

\begin{center}
\resizebox{.8\textwidth}{!}{%
\begin{threeparttable}
\caption{Superconducting gap size of A$_x$Fe$_{2-y}$Se$_2$}
\begin{tabular}{L{0.2\columnwidth}C{0.1\columnwidth}C{0.12\columnwidth}C{0.2\columnwidth}C{0.12\columnwidth}C{0.12\columnwidth}r}
\hline \hline
Sample                                      &T$_C$(K)   &$\Delta_{(\gamma)}$(meV)    &2$\Delta_{(\gamma)}$/K$_B$T$_C$(meV)       &$\Delta_\beta$(meV)   &$\Delta_\alpha$(meV)     &Ref. \\ \hline
K$_{0.8}$Fe$_2$Se$_2$               &31.7          &10.3            &8            &                &7               &\cite{YZhang}   \\
(Tl,Rb)Fe$_{1.72}$Se$_2$            &32            &12                &9            &15            &0               &\cite{DMou}   \\
(Tl,K)Fe$_{1.78}$Se$_2$             &29             &8.5              &7             &                &                 &\cite{XWang}   \\
K$_{0.68}$Fe$_{1.79}$Se$_2$      &32            & 9                &7             &              &                 &\cite{LZhao}   \\
(Tl,K)Fe$_{1.84}$Se$_2$             &28             &8                  &7             &                &                 &\cite{LZhao}   \\
(Tl,Rb)Fe$_{1.72}$Se$_2$            &32            &9.7             &7.5          &8                &0                 &\cite{DMouUn}   \\
KFe$_2$Se$_2$                           &$\sim$28\tnote{*}    &4/1            &3.5/0.95      &               &                &\cite{WLiSTM}   \\
K$_{0.73}$Fe$_{1.67}$Se$_2$     &32             &7                 &5           &                    &                 &\cite{PCai}   \\
\hline \hline
\end{tabular}
\begin{tablenotes}\small
\item[*] T$_C$ is estimated from temperature dependent STS data.
\end{tablenotes}
\end{threeparttable}}%
\end{center}

\clearpage

\begin{figure}[tbp]
\begin{center}
\includegraphics[width=1.0\columnwidth,angle=0]{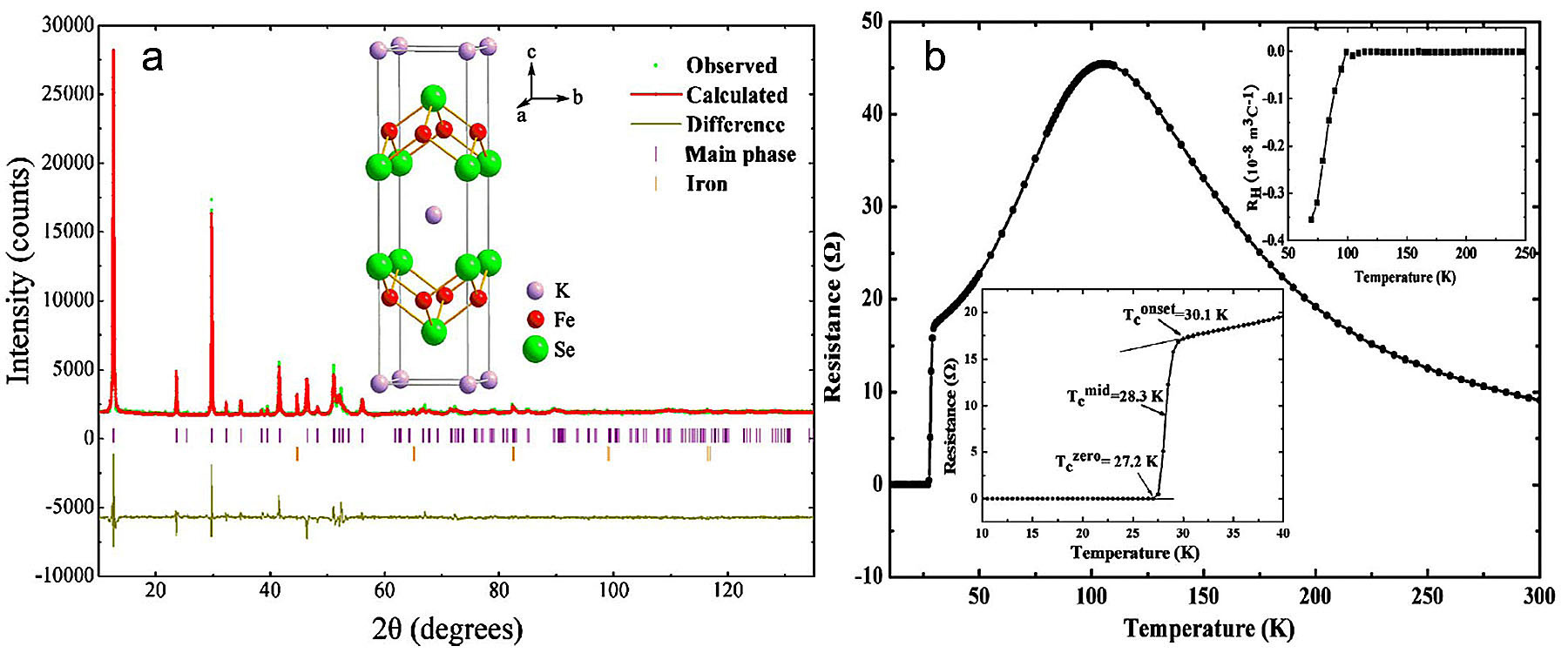}
\end{center}
\caption{(a)Crystal structure of K$_x$Fe$_2$Se$_2$ and its powder X-ray diffraction pattern. (b)Temperature dependent resistance showing a superconducting transition at 30 K. Reprinted from \cite{JGuo}.}
\end{figure}

\begin{figure}[tbp]
\begin{center}
\includegraphics[width=1.0\columnwidth,angle=0]{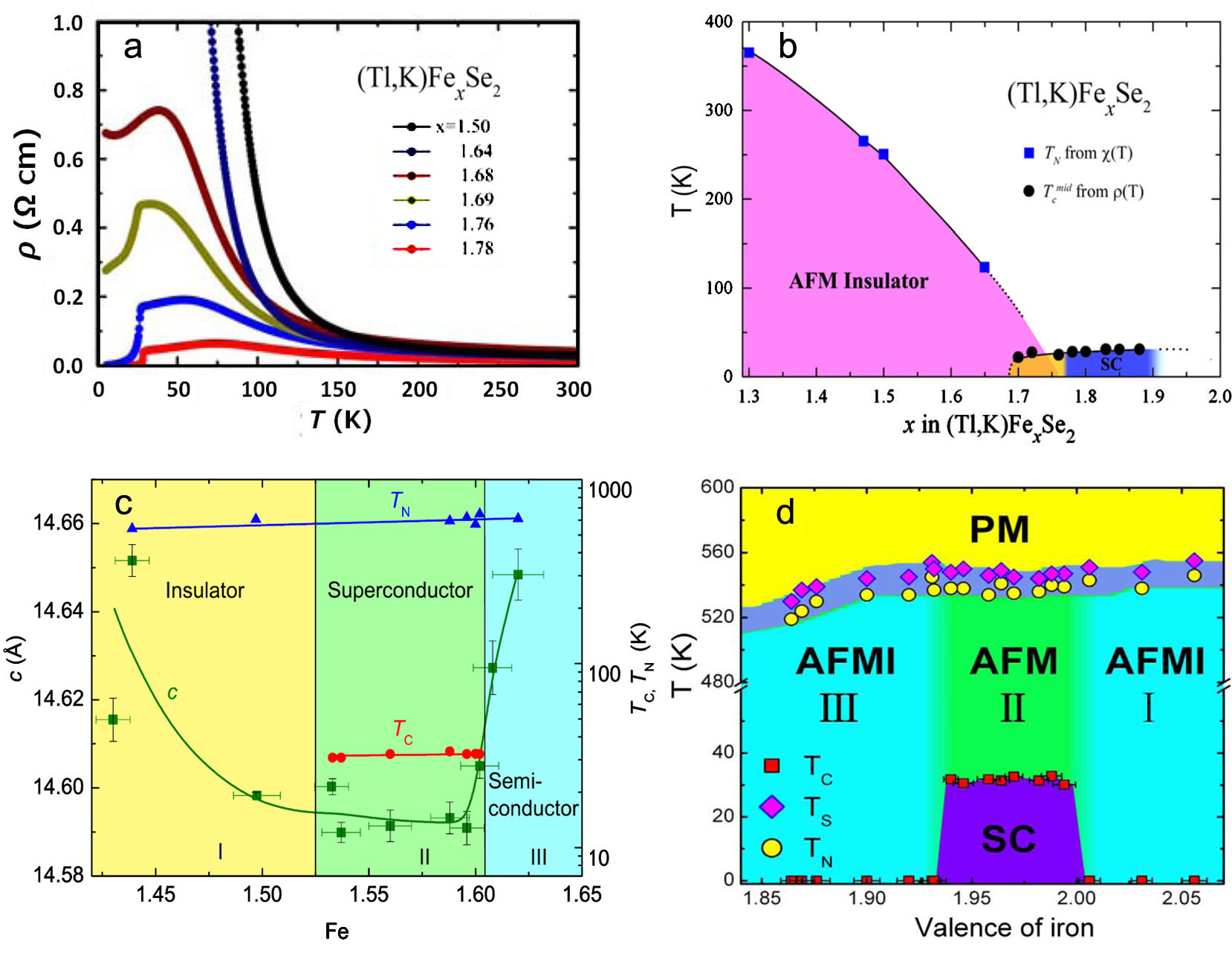}
\end{center}
\caption{(a) Temperature dependent $ab$ plane resistivity of  (Tl$_{1-y}$K$_{y}$)Fe$_x$Se$_2$ with various iron content $x$. (b)Phase diagram of (Tl$_{1-y}$K$_{y}$)Fe$_x$Se$_2$ based on resistivity measurements. (a),(b) are reprinted from \cite{MHFang}. Note that the Antiferromagnietc transition temperature determined by neutron scattering later on is in fact higher than the transition temperature marked in (b).  (c)Phase diagram of Rb$_x$Fe$_{2-y}$Se$_2$ system that shows variation of the lattice constant c, superconducting transition temperature T$_c$ and the N$\acute{e}$el temperature T$_N$ with Fe content. Reprinted from \cite{VTsurkan}. (d)Phase diagram of K$_x$Fe$_{2-y}$Se$_2$ that shows structure, magnetism and superconducting transitions as a function of Fe valence. Reprinted from \cite{YJYan}.}
\end{figure}

\begin{figure}[tbp]
\begin{center}
\includegraphics[width=1.0\columnwidth,angle=0]{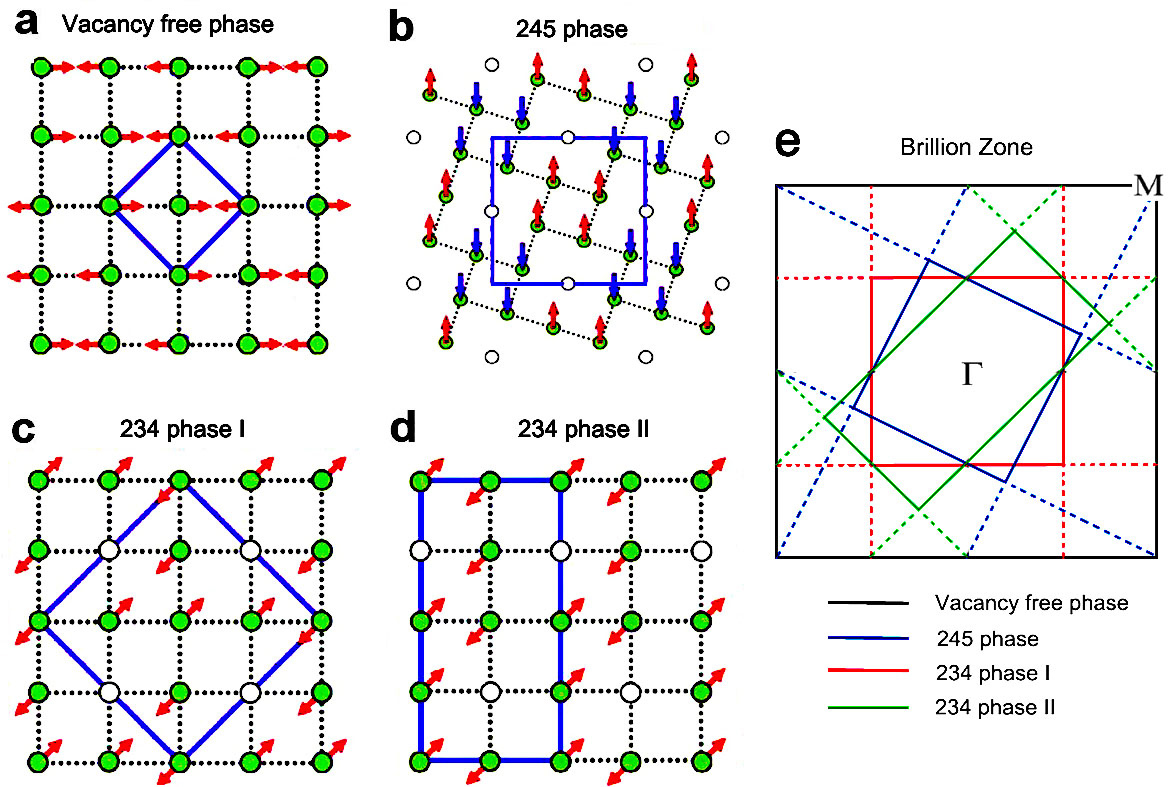}
\end{center}
\caption{Iron lattice and ground state magnetic structures of various vacancy-ordered phases. Blue lines mark corresponding unit cells. (a)Vacancy free phase with calculated bicollinear AFM magnetic structure\cite{XYYanPRBVF}. (b)245 Phase($\sqrt{5}\times\sqrt{5}$ superstructure) with calculated block AFM magnetic structure\cite{CCaoPRL,XYYanPRBVO,RYuPRB,WLv}. It is revealed by Neutron scattering measurements\cite{WBaoCPL,VYuPomjakushinJPCM,VYuPomjakushinPRB}. (c)234 Phase I($2\times2$ superstructure) with calculated c-AFM magnetic structure\cite{XYYanPRL}. (d)234 Phase II($\sqrt{2}\times2\sqrt{2}$ superstructure) with A-collinear AFM magnetic structure\cite{XYYanPRL}. (e)The corresponding Brillion Zones of these structures.}
\end{figure}

\begin{figure}[tbp]
\begin{center}
\includegraphics[width=1.0\columnwidth,angle=0]{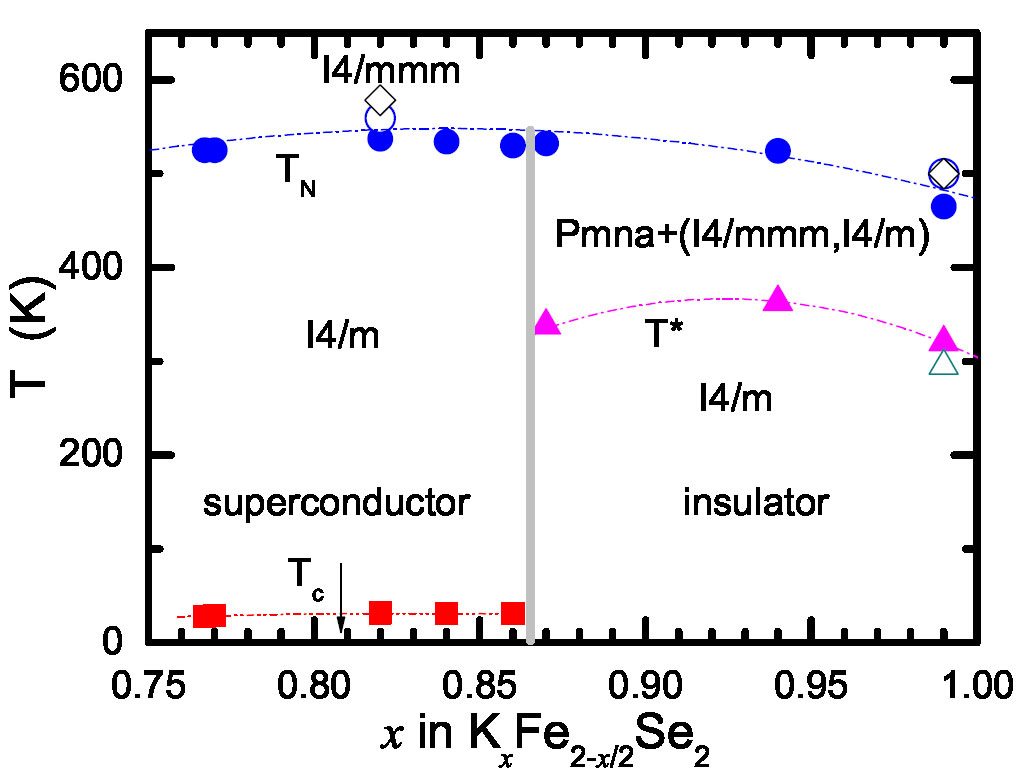}
\end{center}
\caption{Phase diagram of K$_x$Fe$_{2-x/2}$Se$_2$ based on crystal structures measured by temperature dependent Neutron scattering. I4/mmm: vacancy free(Fig.5(a)), I4/m: $\sqrt{5}\times\sqrt{5}$ superstructure(Fig.3b),  Pmna: $\sqrt{2}\times\sqrt{2}$ superstructure. Reprinted from \cite{WBao}. }
\end{figure}

\begin{figure}[tbp]
\begin{center}
\includegraphics[width=1.0\columnwidth,angle=0]{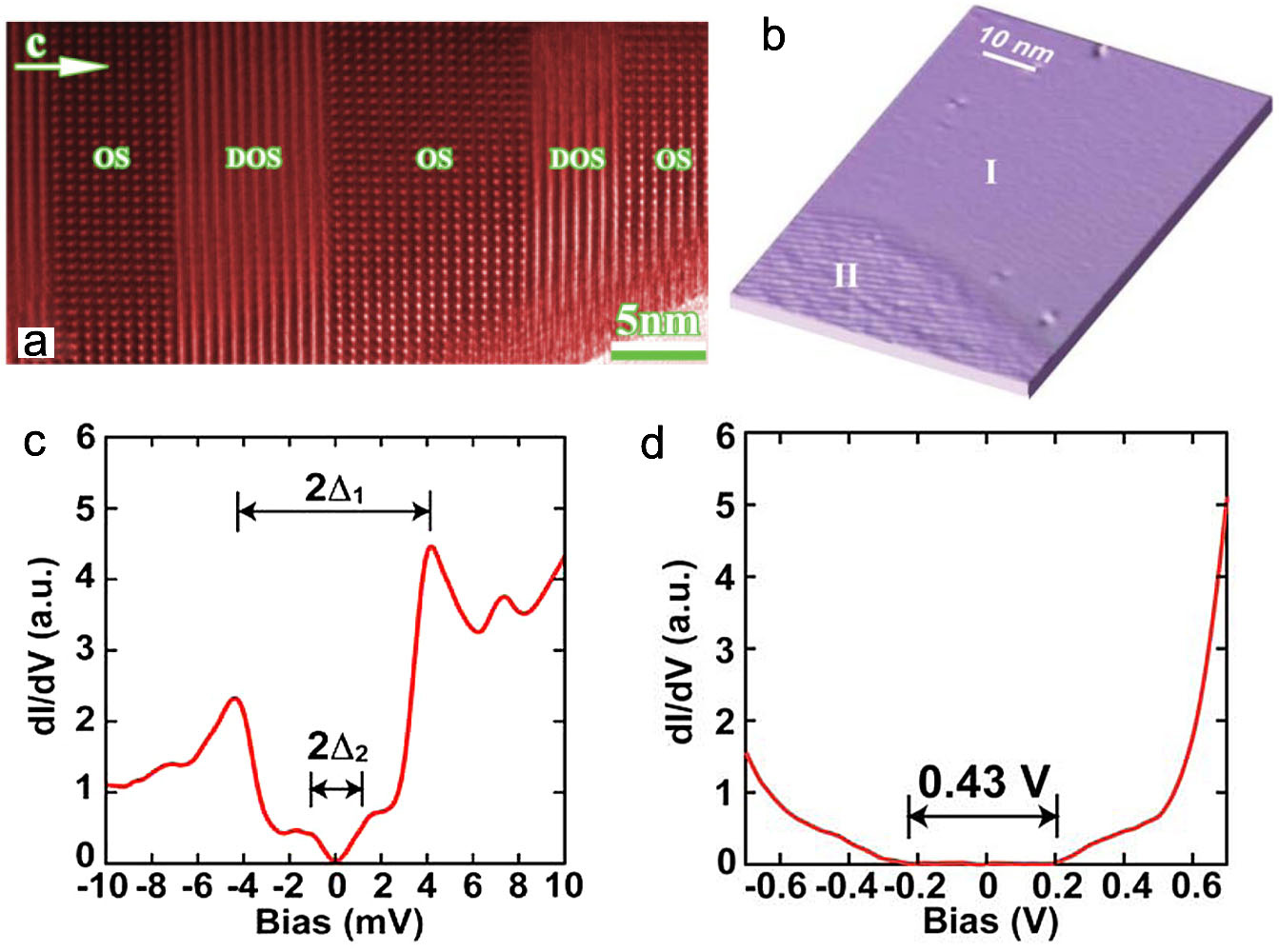}
\end{center}
\caption{(a)TEM results of K$_x$Fe$_{2-y}$Se$_2$ along the [1-30] zone-axis direction. Ordered state (OS) and disordered state(DOS) are separated along the c-axis. Reprinted from \cite{ZWang}. (b)STM topographic image of K$_x$Fe$_{2-y}$Se$_2$ film. Two distinct regions are labeled by I(vacancy free) and II($\sqrt{5}\times\sqrt{5}$ superstructure). (c)Differential conductance spectrum in region I measured at 0.4 K. (d)Differential conductance spectrum in region II. (b)-(d) are reprinted from \cite{WLiSTM}.}
\end{figure}

\begin{figure}[tbp]
\begin{center}
\includegraphics[width=1.0\columnwidth,angle=0]{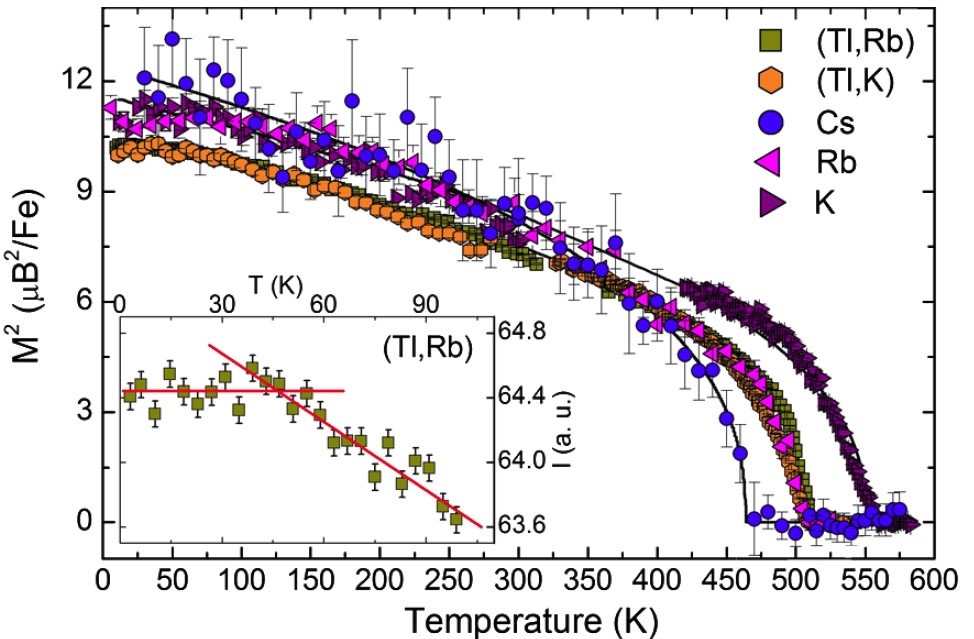}
\end{center}
\caption{Normalized magnetic Bragg intensity that shows the squared magnetic moment in block AFM as a function of temperature. Inset: Magnetic (101) peak of (Tl, Rb)$_2$Fe$_4$Se$_5$ showing a transition around $T_C$. Reprinted from \cite{FYe}.}
\end{figure}

\begin{figure}[tbp]
\begin{center}
\includegraphics[width=1.0\columnwidth,angle=0]{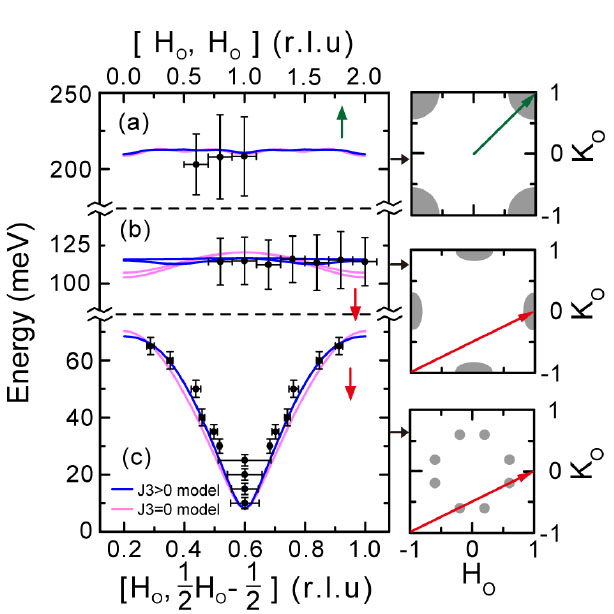}
\end{center}
\caption{Spin-wave dispersions of Rb$_{0.89}$Fe$_{1.58}$Se$_2$ along high-symmetry directions from Inelastic Neutron Scattering measurements. (a) Highest energy optical band; (b)Medium energy optical band; and (c)Acoustic spin wave mode. Corresponding cuts are illustrated at right inserts. Bands are fitted using the Heisenberg Hamiltonian. Reprinted from \cite{MWangIN}.}
\end{figure}

\begin{figure}[tbp]
\begin{center}
\includegraphics[width=1.0\columnwidth,angle=0]{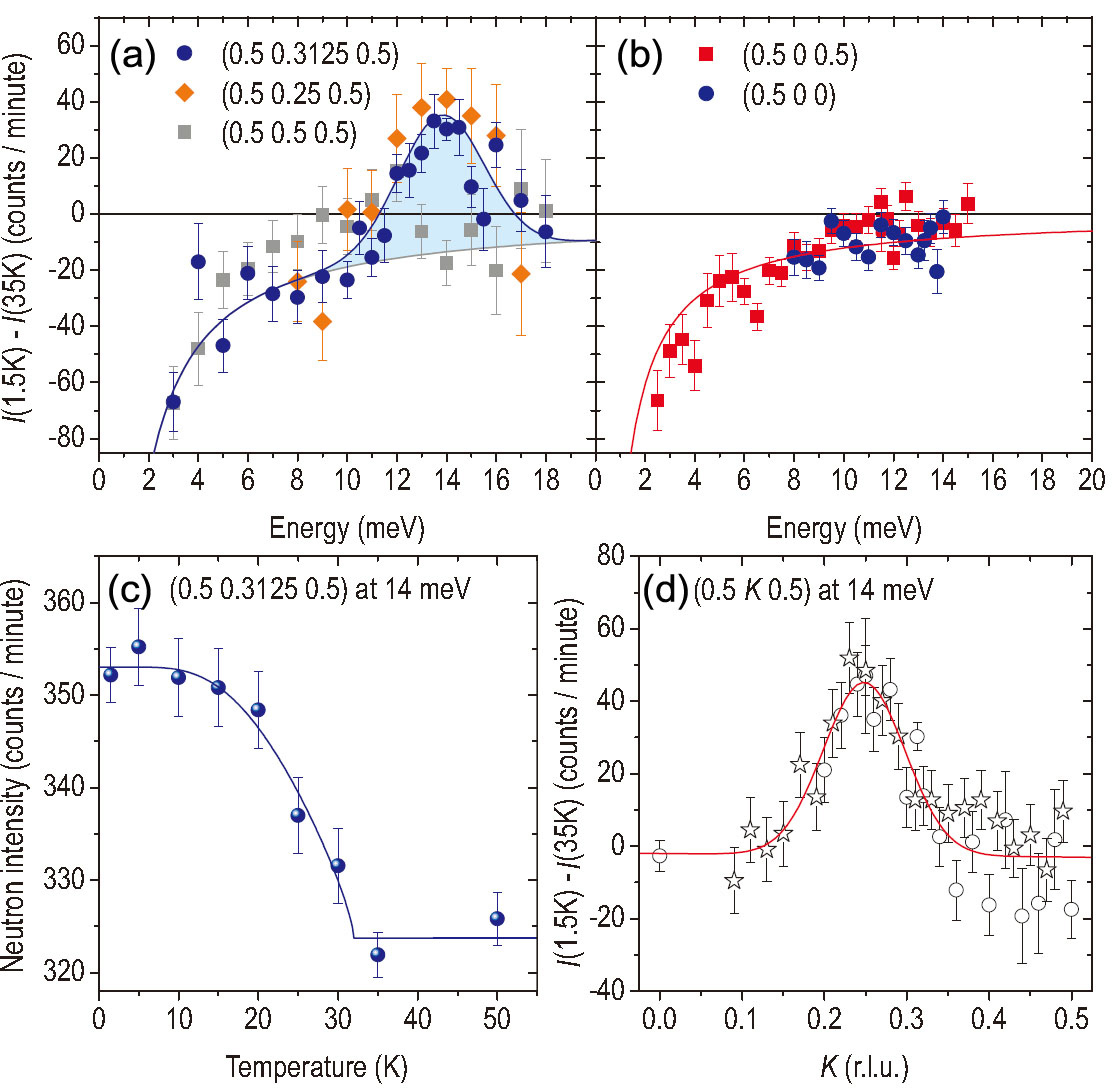}
\end{center}
\caption{Spin resonance mode revealed by Inelastic Neutron Scattering. (a)Intensity difference between the superconducting state and normal state at three Q-vectors. Resonance peak (shaded region) is found around 14 meV both at (0.5, 0.25, 0.5) and (0.5, 0.3125, 0.5) (1 Fe Brillioun Zone). (b)The same plot as (c), but for  (0.5, 0, 0.5) and (0.5, 0, 0), that reveals no resonance. (c)Temperature dependence of the INS intensity at 14 meV and at (0.5, 0.3125, 0.5). It shows an order-parameter-like behavior with an onset at T$_C$. (d)Momentum scans along the BZ boundary of intensity difference between superconducting and normal state, with a maximum at the commensurate wave vector (0.5, 0.25, 0.5). The solid line is a Gaussian fit with a linear background. Reprinted from \cite{JTPark}.}
\end{figure}

\begin{figure}[tbp]
\begin{center}
\includegraphics[width=1.0\columnwidth,angle=0]{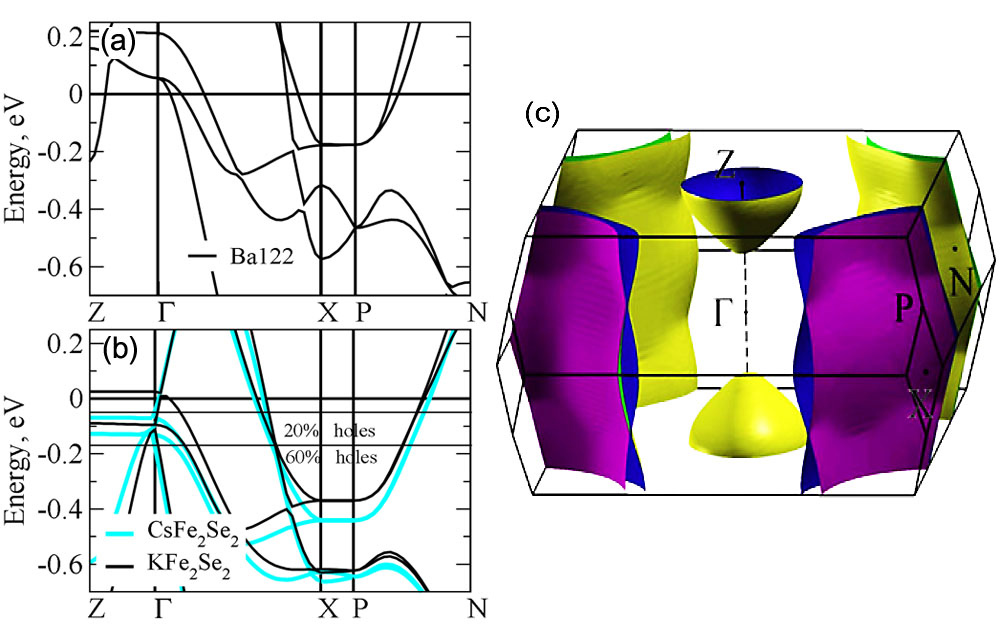}
\end{center}
\caption{(a)LDA calculated band dispersions for BaFe$_2$As$_2$. (b)KFe$_2$Se$_2$ (black lines) and CsFe$_2$Se$_2$ (cyan lines). Fermi level position for 20\% and 60\% hole doping are marked. Reprinted from \cite{IANekrasov}. (c)Calculated Fermi Surface for KFe$_2$Se$_2$. Lattice parameters and the atomic positions are optimized. Reprinted from \cite{IRShein}.}
\end{figure}

\clearpage

\begin{figure}[tbp]
\begin{center}
\includegraphics[width=1\columnwidth,angle=0]{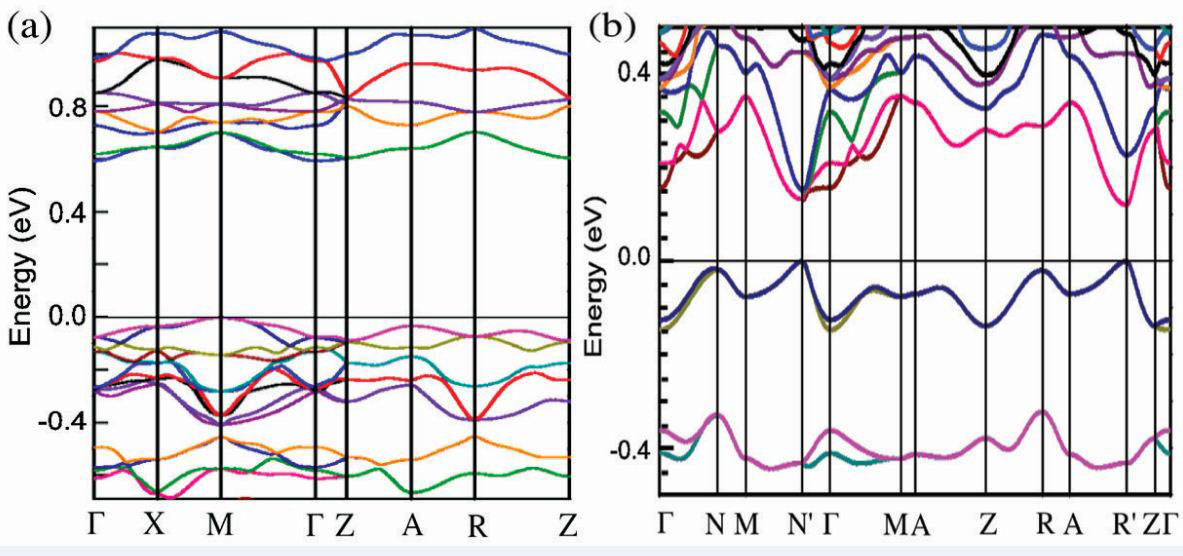}
\end{center}
\caption{Calculated electronic band structure of  Fe vacancy ordered phase. (a)K$_{0.8}$Fe$_{1.6}$Se$_2$(245 phase) in the ground state with a block AFM order(Fig.3b). Reprinted from \cite{XYYanPRBVO}. (b)KFe$_{1.5}$Se$_2$(234 phase II) with an A-collinear AFM order(Fig.3d). Reprinted from \cite{XYYanPRL}.}
\end{figure}

\begin{figure}[tbp]
\begin{center}
\includegraphics[width=1.0\columnwidth,angle=0]{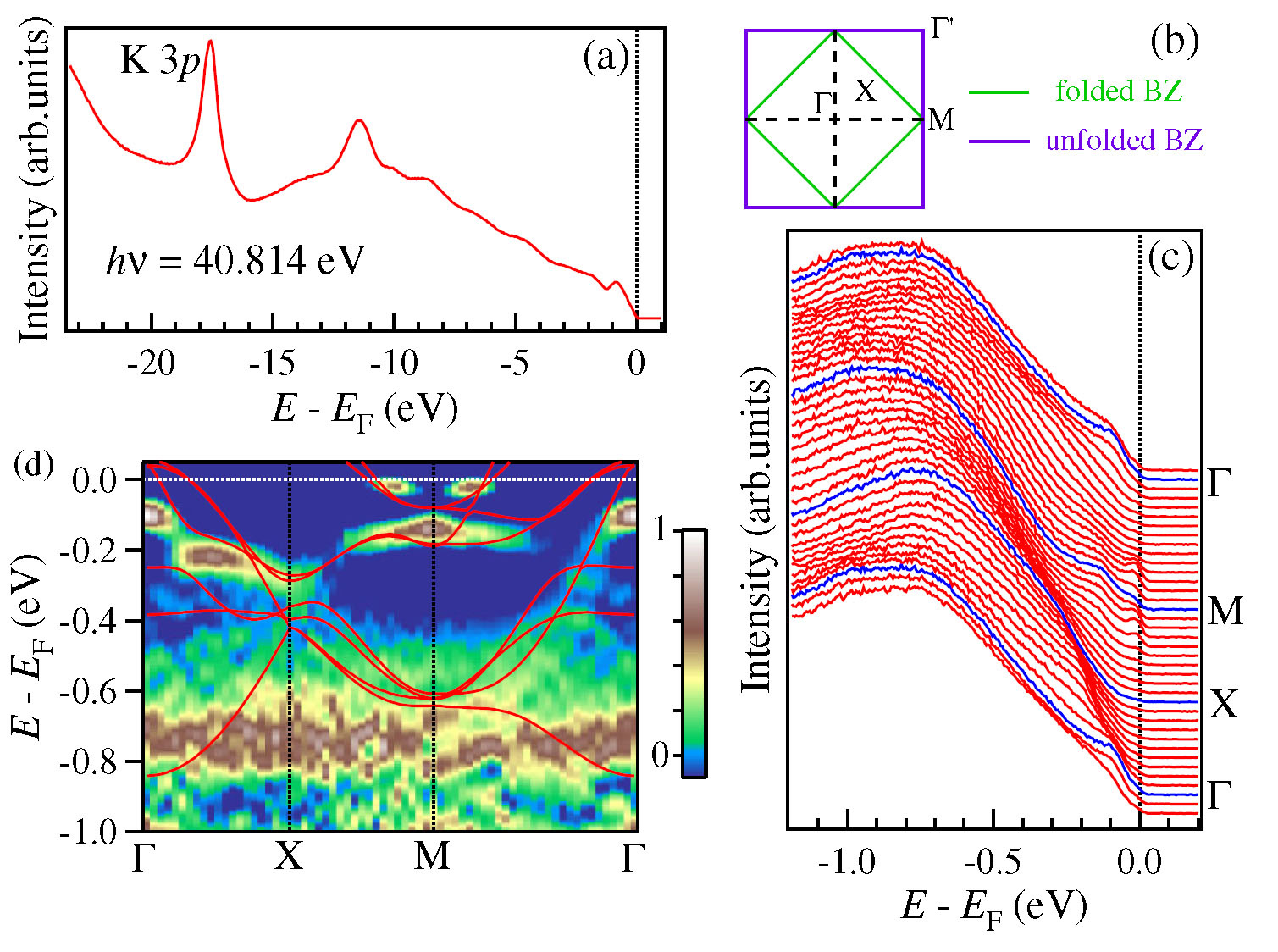}
\end{center}
\caption{(a)Photoemission spectra with a wide energy, integrated along $\Gamma-M$ within $\pm$15$^{\circ}$. (b)Schematic definition of the $\Gamma$(0, 0), $M$($\pi$, 0) and $X$($\pi$/2, $\pi$/2) high symmetry points. (c)EDCs along several high symmetry directions. (d)Second derivative intensity plot along high symmetry lines, together with LDA calculated band structures of KFe$_2$Se$_2$($K_z$=0), which have been shifted up by 170 meV to account for the electron doping and then renormalized by a factor 2.5. Reprinted from \cite{TQian}.}
\end{figure}

\begin{figure}[tbp]
\begin{center}
\includegraphics[width=1.0\columnwidth,angle=0]{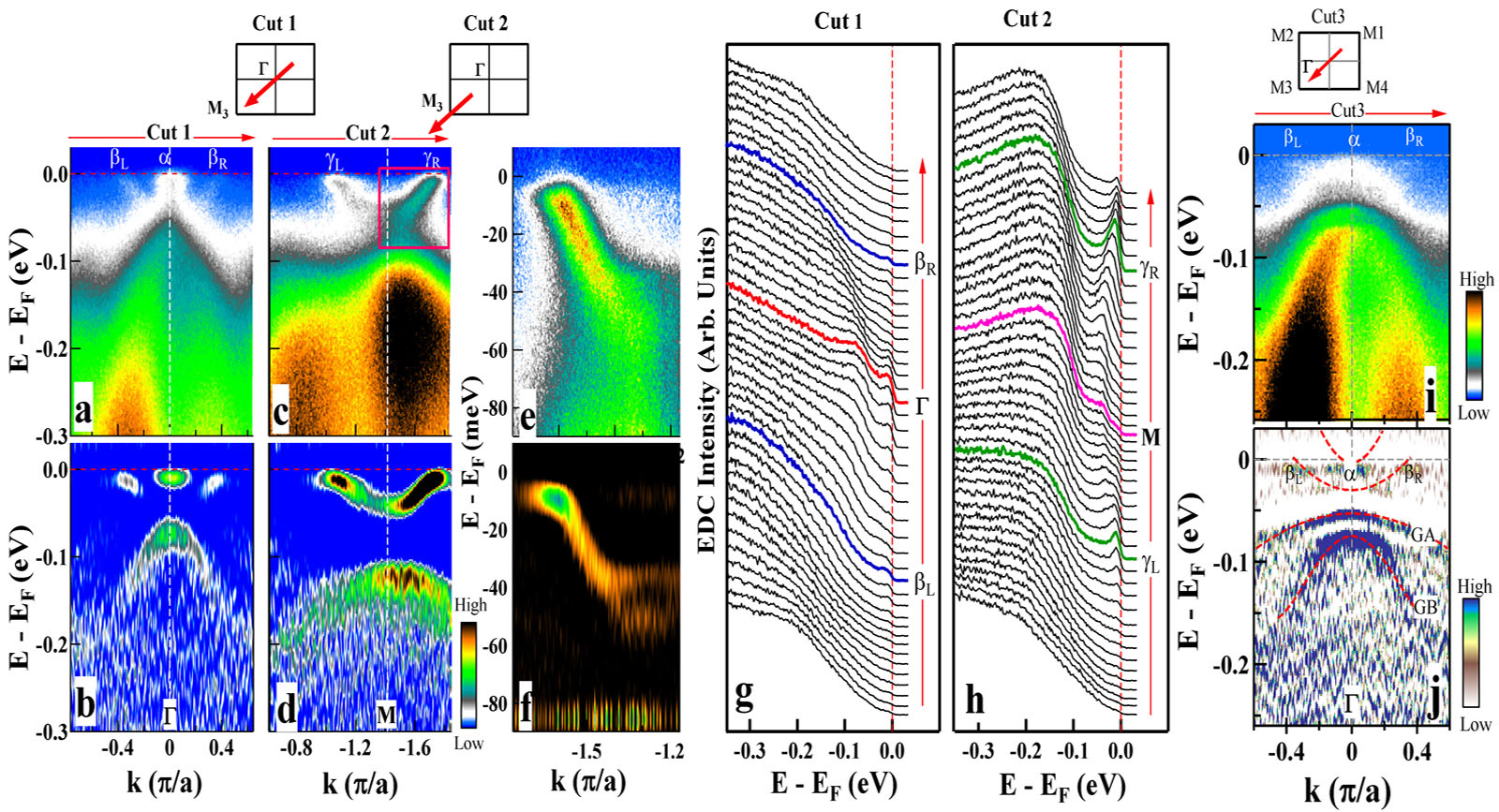}
\end{center}
\caption{Band structure and photoemission spectra of A$_x$Fe$_{2-y}$Se$_2$ superconductors measured along two high symmetry cuts\cite{DMou,LZhao,DMouUn}. Cut locations are illustrated at the top-left and top-right inserts. (a)(c)Measured band structure images of (Tl$_{0.58}$Rb$_{0.42})$Fe$_{1.72}$Se$_2$ along cut1 and cut2, respectively. (b)(d)Their corresponding EDC second derivative images. (e)(f)Fine measurement of the band structure in red square of (c) and its corresponding EDC second derivative images. (g)(h)Photoemission spectra (EDCs) corresponding to (a) and (c), respectively. EDCs of Fermi crossing and high symmetry points are marked. (i)Measured band structure of K$_{0.68}$Fe$_{1.79}$Se$_2$ along Cut3. (j)Corresponding EDC second derivative images. Besides two electron bands, two hole bands are observed below Fermi level with different photon energies around $\Gamma$, denoted as GA and GB.}
\end{figure}

\begin{figure}[tbp]
\begin{center}
\includegraphics[width=.8\columnwidth,angle=0]{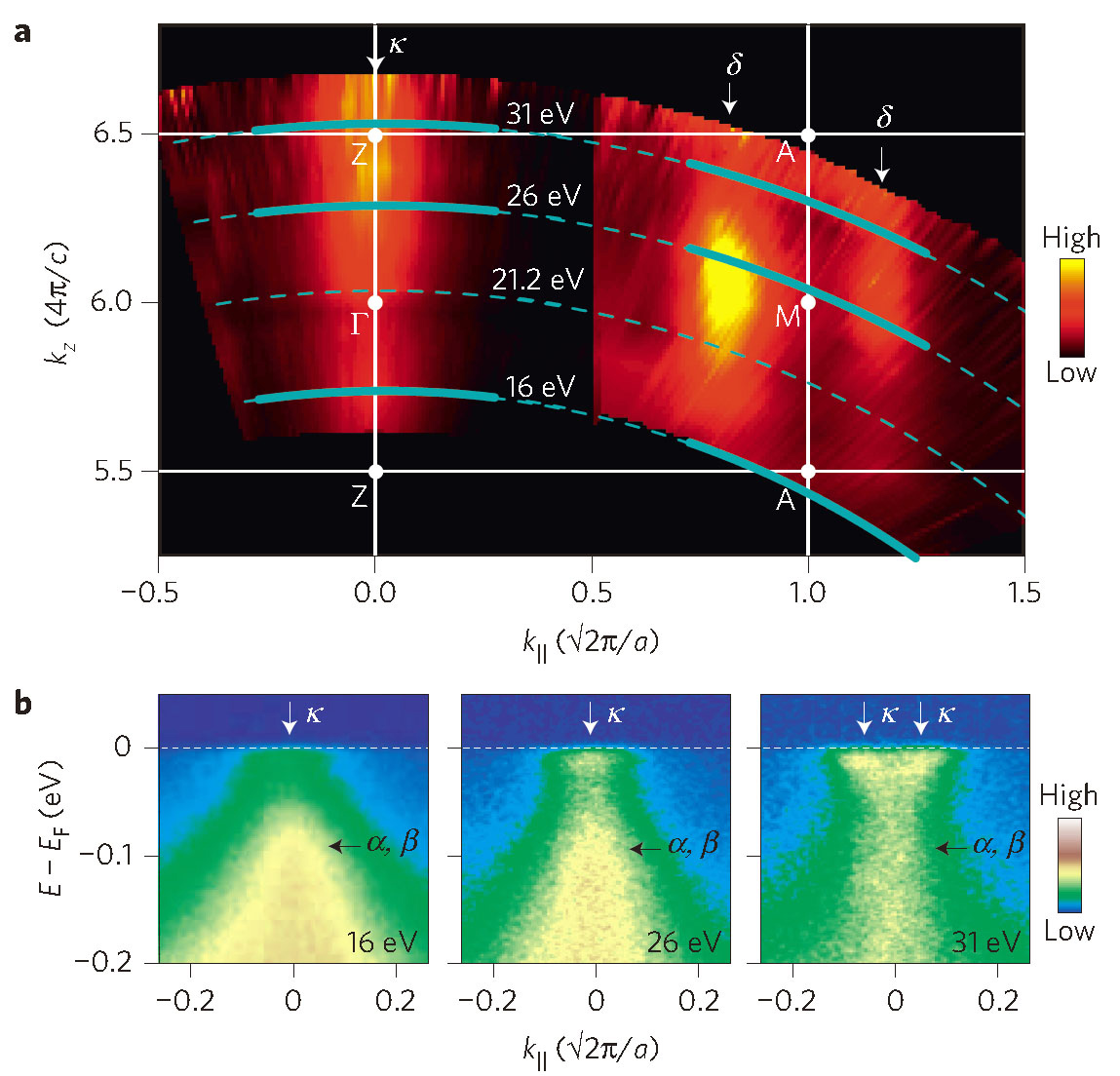}
\end{center}
\caption{The Fermi surface and band structure as a function of k$_z$ in K$_{0.8}$Fe$_2$Se$_2$\cite{YZhang}. (a)The photoemission intensity in the $\Gamma$ZAM plane. (b)Measured band structures around $\Gamma$ with three different photon energy. $k$, $\delta$ correspond to $\alpha$ and $\gamma$ in Fig. 12 respectively.}
\end{figure}

\begin{figure}[tbp]
\begin{center}
\includegraphics[width=.9\columnwidth,angle=0]{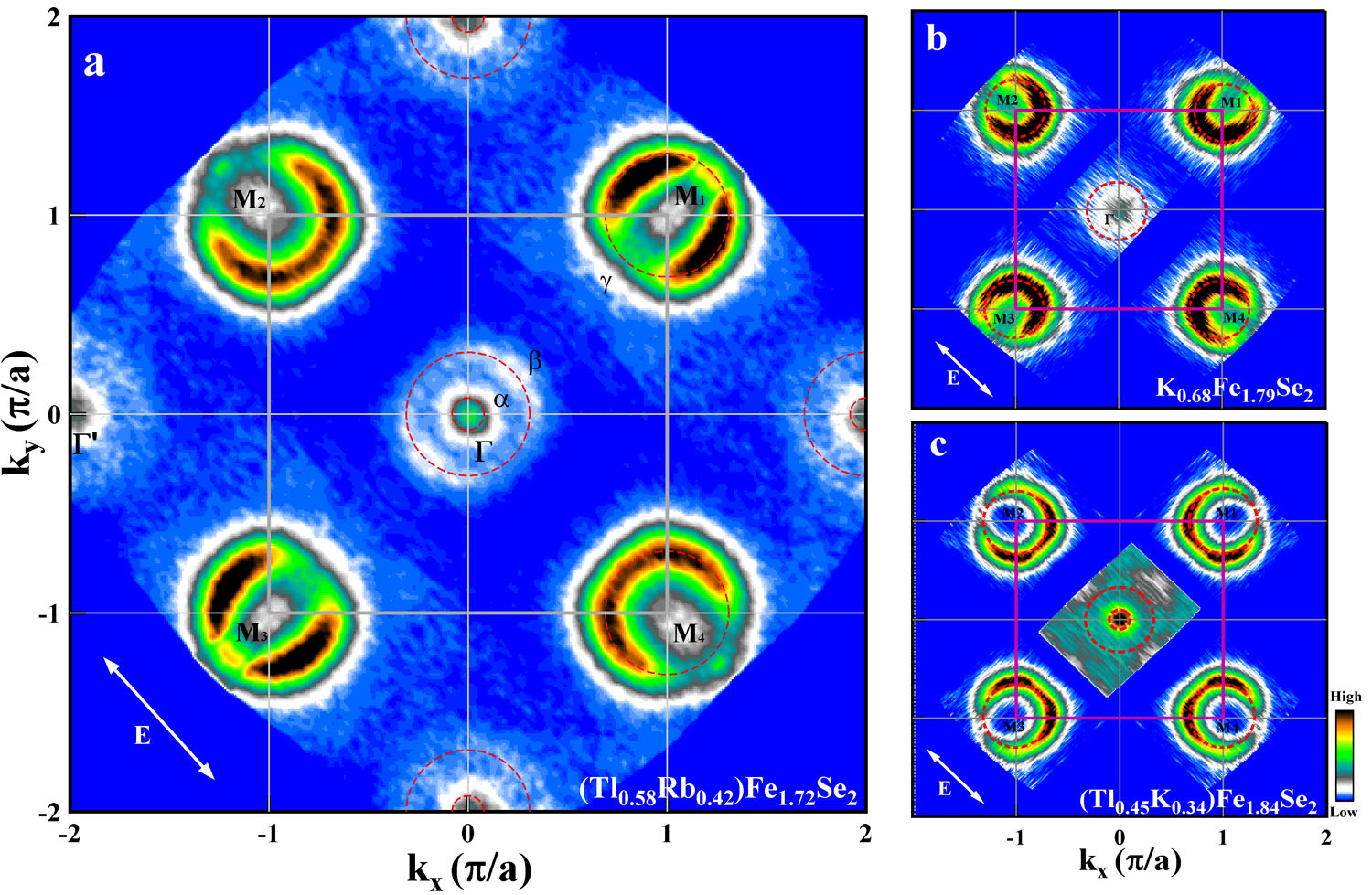}
\end{center}
\caption{Measured Fermi Surface of three A$_x$Fe$_{2-y}$Se$_2$ superconductors. (a)Fermi Surface of (Tl$_{0.58}$Rb$_{0.42})$Fe$_{1.72}$Se$_2$. (b)K$_{0.68}$Fe$_{1.79}$Se$_2$. (c) (Tl$_{0.45}$K$_{0.34}$)Fe$_{1.84}$Se$_2$. Two Fermi surface sheets are observed around the $\Gamma$ point which are marked as $\alpha$ for the inner small sheet and $\beta$ for the outer large one. One Fermi Surface sheet around M is marked as $\gamma$. (a) reprinted from \cite{DMou}, (b)(c) reprinted from \cite{LZhao}.}
\end{figure}

\begin{figure}[tbp]
\begin{center}
\includegraphics[width=1.0\columnwidth,angle=0]{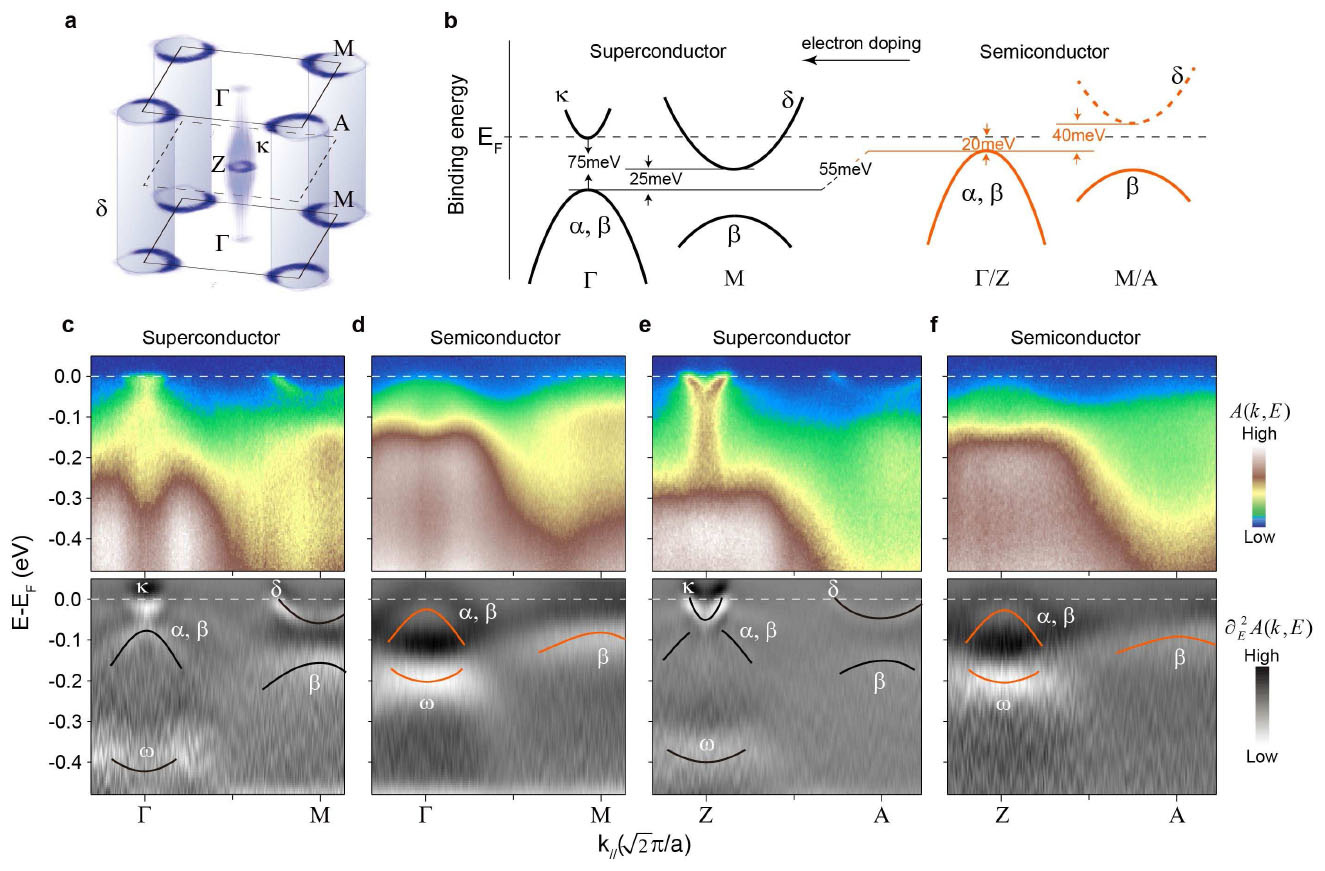}
\end{center}
\caption{Low energy electronic structures of the superconducting and semiconducting K$_x$Fe$_{2-y}$Se$_2$. (a)The Fermi surface of the superconducting phase in the three-dimensional Brillouin zone. (b)The sketch of the band structure evolution from the semiconductor to the
superconductor. (c)The photoemission intensities (upper panel) and its second derivative with respect to energy (lower panel) along the $\Gamma$ - $M$ direction for the superconductor taken at normal state. (d) Same as (c) but for semiconductor. (e)(f)Same as (c)(d) along $Z$ - $A$ direction. Reprinted from \cite{FChen}.}
\end{figure}

\begin{figure}[tbp]
\begin{center}
\includegraphics[width=.9\columnwidth,angle=0]{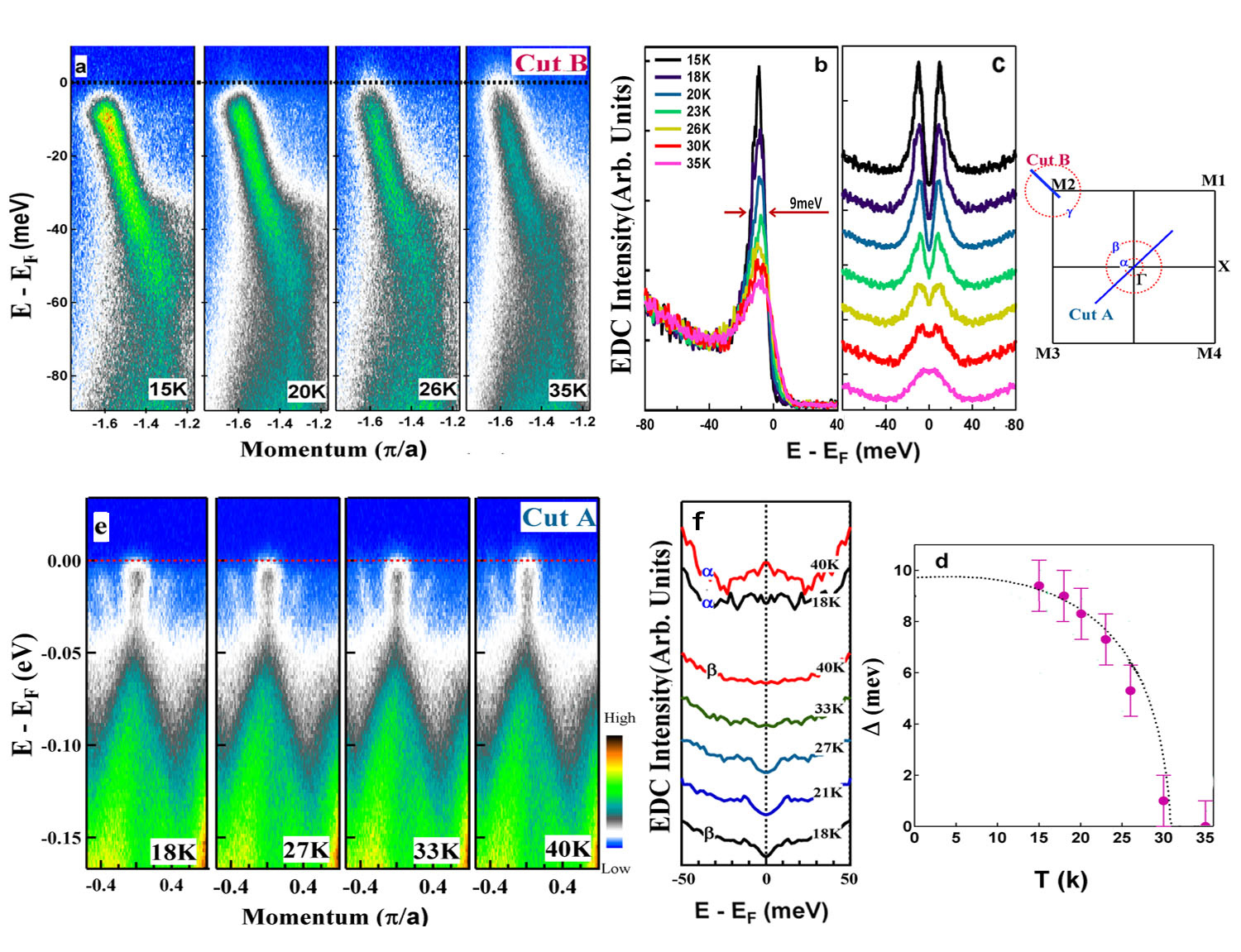}
\end{center}
\caption{Temperature dependence of band structures and superconducting gap of (Tl$_{0.58}$Rb$_{0.42})$Fe$_{1.72}$Se$_2$\cite{DMou,DMouUn}. Cut locations are illustrated at top-right inset. (a)Temperature dependent band structure image plots of (Tl$_{0.58}$Rb$_{0.42})$Fe$_{1.72}$Se$_2$ along cut B around $M$ point. (b) EDCs at the Fermi crossing extracted from (a) at different temperatures. (c)Symmetrized EDCs to remove the  Fermi-Dirac function and visualize gap opening. (d)Temperature dependent superconducting gap of (Tl$_{0.58}$Rb$_{0.42})$Fe$_{1.72}$Se$_2$ on the $\gamma$ FS. The gap size is obtained by picking up the peak position in (c). Dashed line is a BCS-form fit with $\Delta_0$=9.7meV. (e)Temperature dependent band images along cut A around $\Gamma$ point. (f)Symmetrized EDCs of the $\beta$ band crossing point and $\Gamma$ point.}
\end{figure}

\begin{figure}[tbp]
\begin{center}
\includegraphics[width=1.0\columnwidth,angle=0]{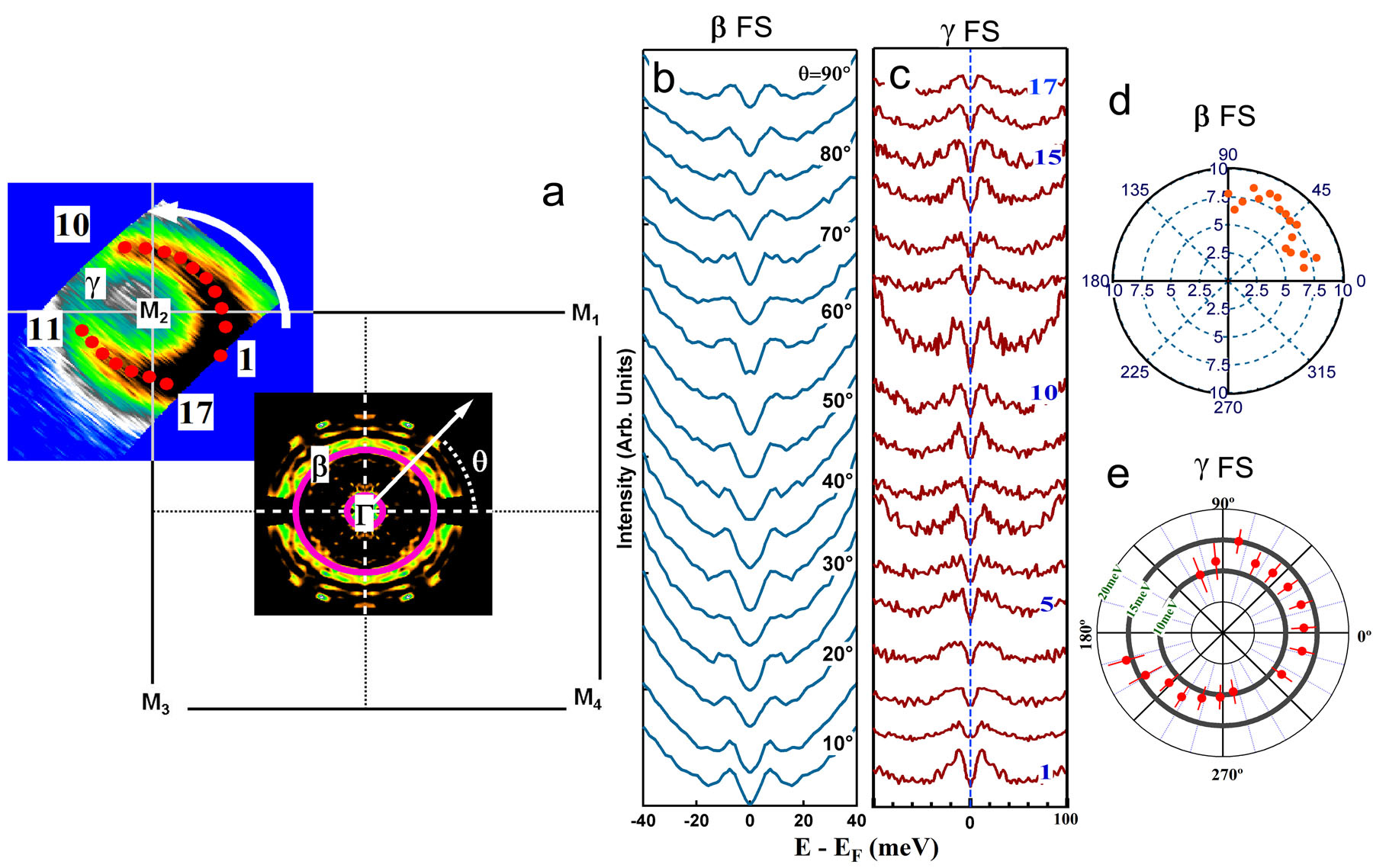}
\end{center}
\caption{Momentum dependent superconducting gap of (Tl$_{0.58}$Rb$_{0.42})$Fe$_{1.72}$Se$_2$\cite{DMou,DMouUn}. (a)Fermi surface mapping around M$_2$ at 15 K\cite{DMou} and a second derivative image of measured Fermi Surface around $\Gamma$ at 18 K measured by high resolution laser-based ARPES\cite{DMouUn}. (b)Symmetrized EDCs at the Fermi crossing of $\beta$ Fermi surface sheet; The angle is defined in (a). (c)Symmetrized EDCs at the Fermi crossing of $\gamma$ FS sheet as marked in (a). (d)(e)Momentum dependent superconducting gap size for the $\beta$ Fermi surface sheet around $\Gamma$ and $\gamma$ Fermi surface sheet around M, respectively.}
\end{figure}

\begin{figure}[tbp]
\begin{center}
\includegraphics[width=.9\columnwidth,angle=0]{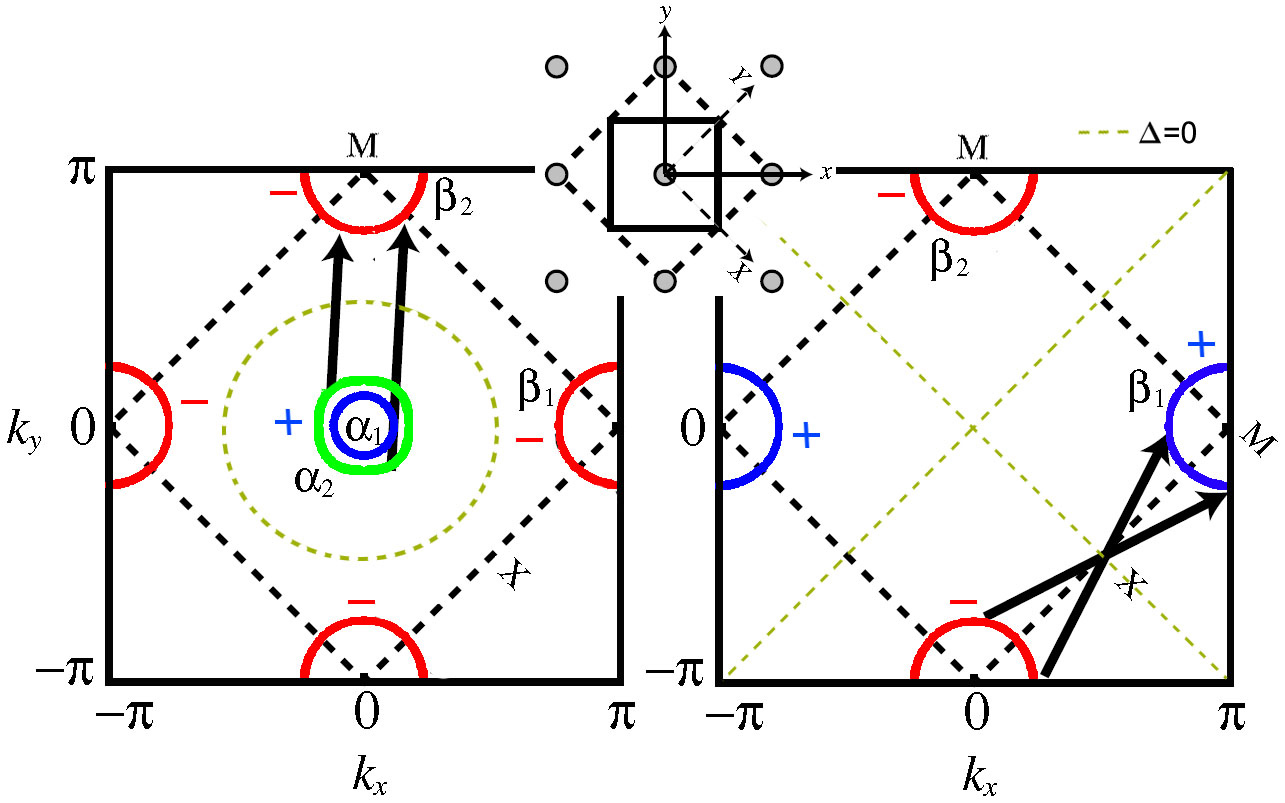}
\end{center}
\caption{Schematic gap symmetry based on electron scattering\cite{KKuroki}. Left: s$\pm$ gap structure in iron-pnictide superconductor, due to electron scattering between the hole pockets in the BZ center and the electron pocket in the BZ corner. Right: Nodeless d-wave gap symmetry as a result of electron scattering between two electron pockets around BZ corner. Gap nodes are marked by yellow dashed lines.}
\end{figure}

\begin{figure}[tbp]
\begin{center}
\includegraphics[width=.9\columnwidth,angle=0]{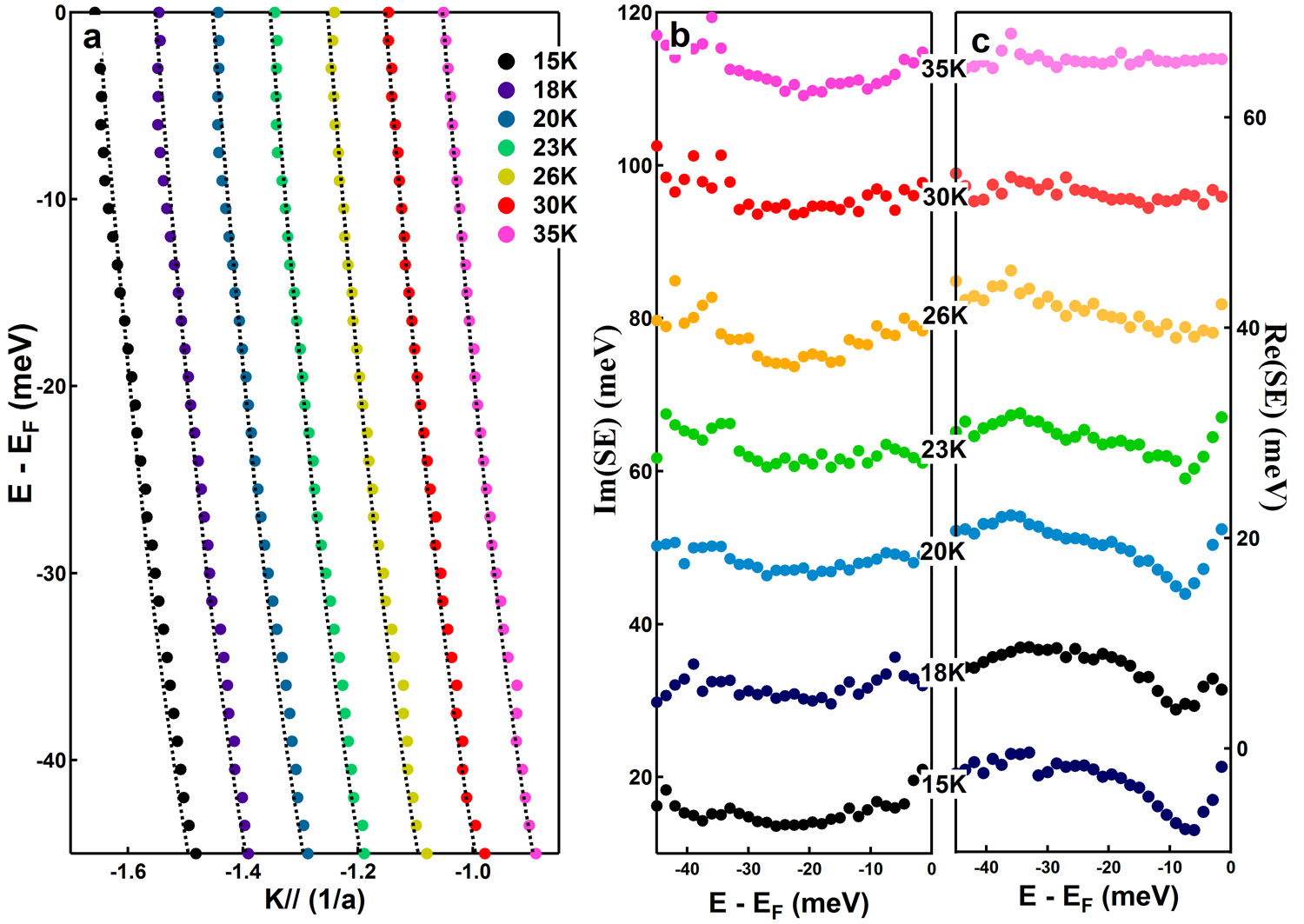}
\end{center}
\caption{Temperature dependent electronic dynamics around $M$ point\cite{DMouUn}. (a)Fitted band dispersions at different temperatures along cut B in Fig. 16. Black dashed lines are selected parabolic ``bare band". (b)(c)Extracted imaginary and real parts of electron self-energy.}
\end{figure}

\end{document}